\documentclass[numsec,webpdf,modern,large,namedate]{oup-authoring-template}

\onecolumn

\usepackage[doublespacing]{setspace}
\usepackage[fontsize=12pt]{fontsize}

\RequirePackage{amsthm,amsmath,amsfonts,amssymb}
\RequirePackage[authoryear]{natbib}
\RequirePackage{graphicx}
\usepackage{subfigure}

\usepackage{float}
\usepackage{multirow}
\usepackage{diagbox}
\usepackage[figuresright]{rotating}
\usepackage{pdflscape}
\usepackage{bm}
\usepackage{algorithm}
\usepackage{algpseudocode}
\usepackage{enumerate}
\usepackage{booktabs}
\usepackage{mathabx} 

\renewcommand\figurename{Figure}

\newtheorem{thm}{Theorem}

\newtheorem{assump}{Assumption}

\newtheorem{emp}{Example}

\usepackage{xr-hyper}
\usepackage{hyperref}
\makeatletter
\newcommand*{\addFileDependency}[1]{
\typeout{(#1)}
\@addtofilelist{#1}
\IfFileExists{#1}{}{\typeout{No file #1.}}
}
\makeatother
\newcommand{\relu}{\bm{\sigma}}

\newcounter{cnstcnt}

\newcounter{bnstcnt}

\newcounter{dbnstcnt}


\makeatletter
\newcommand*{\centerfloat}{%
  \parindent \z@
  \leftskip \z@ \@plus 1fil \@minus \marginparwidth
  \rightskip \leftskip
  \parfillskip \z@skip}
\makeatother

\usepackage{enumerate}

\begin{document}

\journaltitle{Journals of the Royal Statistical Society}
\DOI{DOI HERE}
\copyrightyear{XXXX}
\pubyear{XXXX}
\access{Advance Access Publication Date: Day Month Year}
\appnotes{Original article}

\firstpage{1}

\title[Deep SemiPDE Models]{Deep Semiparametric Partial Differential Equation Models}

\author[1]{Ziyuan Chen}
\author[1]{Shunxing Yan}
\author[1,$\ast$]{Fang Yao}

\authormark{Ziyuan Chen, Shunxing Yan and Fang Yao}

\address[1]{\orgdiv{Department of Probability and Statistics, School of Mathematical Sciences, Center for Statistical Science}, 
\orgname{Peking University}, \orgaddress{ \state{Beijing}, \country{China}}}

\corresp[$\ast$]{Correspondence: Fang Yao, Email: \href{fyao@math.pku.edu.cn}{fyao@math.pku.edu.cn}}

\received{Date}{0}{Year}
\revised{Date}{0}{Year}
\accepted{Date}{0}{Year}



\abstract{In many scientific fields, the generation and evolution of data are governed by partial differential equations (PDEs) which are typically informed by established physical laws at the macroscopic level to describe general and predictable dynamics. However, some complex influences may not be fully captured by these laws at the microscopic level due to limited scientific understanding. This work proposes a unified framework to model, estimate, and infer the mechanisms underlying data dynamics. We introduce a general semiparametric PDE (SemiPDE) model that combines interpretable mechanisms based on physical laws with flexible data-driven components to account for unknown effects. The physical mechanisms enhance the SemiPDE model's stability and interpretability, while the data-driven components improve adaptivity to complex real-world scenarios. A deep profiling M-estimation approach is proposed to decouple the solutions of PDEs in the estimation procedure, leveraging both the accuracy of numerical methods for solving PDEs and the expressive power of neural networks. For the first time, we establish a semiparametric inference method and theory for deep M-estimation, considering both training dynamics and complex PDE models. We analyze how the PDE structure affects the convergence rate of the nonparametric estimator, and consequently, the parametric efficiency and inference procedure enable the identification of interpretable mechanisms governing data dynamics. Simulated and real-world examples demonstrate the effectiveness of the proposed  methodology and support the theoretical findings.}
\keywords{partial differential equation, semiparametric efficiency, neural networks, penalized M-estimation}

\maketitle
\section{Introduction}
\label{intro}
In many scientific fields, the dynamic processes of data generation and evolution are often described by Partial differential equations (PDEs). They encompass macroscopic phenomena governed by the heat, wave, and Navier-Stokes equations, such as marine environments, the water vapor cycle, and atmospheric changes \citep{stockie2011mathematics, rasmussen2012dynamic, jeng2013integrated}. Similarly, PDEs such as the Reaction-Diffusion and FitzHugh-Nagumo equations model microscopic processes, including neurobiology and electrochemical reactions \citep{sgura2019parameter, singh2022analysis}. A central research interest is to formulate dynamic processes as PDEs by physical laws and analyze the relationships between PDEs and their solutions. 
In computational mathematics, the forward problem of PDEs which involves numerically solving specified PDEs, has been addressed using various methods, including the finite element method \citep[e.g.][]{johnson2009numerical} and machine learning approaches \citep{raissi2019physics, lu2021deepxde}. These methods are then employed to simulate state variables in fields such as computational fluid dynamics and structural mechanics \citep{pantano2003nonlinear,blazek2015computational}.

In this paper, we consider the statistical modeling of data dynamics governed by PDEs. The data $(X_i,Y_i),1\le i\le n$ are observed according to the model $Y_i = u(X_i) +\varepsilon_i,$ where $u(x)$ is the solution to PDEs governing the data dynamics, and $\varepsilon_i$ are observation errors. Our goal is to estimate and infer the PDEs from noisy observations using statistical learning methods. This approach is important for analyzing and explaining key variables, phenomena, and governing laws in the data dynamics. While related to inverse problems for PDEs, which typically assume known equations and focus on recovering unknown inputs such as coefficients, sources, or boundary conditions \citep{isakov2006inverse}, our work emphasizes learning the PDEs themselves from noisy data.

Various statistical methods have been developed to address estimation problems in which the governing equations have explicit functional forms but contain a finite number of unknown parameters. 
Early work in this area dates back to epidemic models, where parameters are estimated in ordinary differential equations (ODEs) which have explicit solutions \citep{ho1995rapid,wu1999population}. Two-step methods \citep{liang2008parameter,tan2024green} involve an initial data pre-smoothing step, followed by model fitting. However, their performances heavily depend on the effectiveness of the pre-smoothing, which does not utilize the specific structure of the equations and is sensitive to noise. For more general ODEs, \cite{ramsay2007parameter} proposed the parameter cascading method based on a generalized smoothing and profiling approach. This method has since been widely applied, leading to several extensions including modifications for ODEs with time-varying parameters \citep{cao2012penalized}, ODEs with mixed-effects \citep{wang2014estimating}, and PDEs \citep{xun2013parameter}. Gaussian processes can be used to estimate and infer parameters in ODEs \citep{yang2021inference}, and Bayesian methods are also employed \citep{nickl2017bayesian,vadeboncoeur2023fully}. 

The above methods are limited to parametric PDE models and may not accurately capture the data-generating processes in realistic and complex environments.
This gap has prompted researchers to explore the unknown forms in PDEs, particularly with the rise of machine learning techniques such as neural networks. \cite{dai2022kernel} introduced the kernel ODEs method for fully nonparametric discovery of the underlying ODE structures. Deep learning  \citep{berg2019data} and reinforcement learning approaches \citep{du2024discover} have been proposed for data-driven PDE discovery. Physics-informed neural networks \citep{raissi2019physics,podina2023universal} and the universal differential equations framework \citep{rackauckas2020universal} have been widely applied to identify unknown PDE structures, aided by symbolic regression \citep{tenachi2023deep,podina2023universal}. Infinite-dimensional Bayesian methods are also used to infer unknown source and force terms in PDEs \citep{nickl2023bayesian,jia2023variational}. These works focus on applying fully nonparametric methods to approximate the governing PDEs. However, such methods are often unstable and lack interpretability.

From the above discussion, it is evident that existing methods often struggle to balance interpretability with flexibility in modeling governing mechanisms. Traditional parameter estimation methods for PDEs focus exclusively on the established physical laws and may be misspecified due to the omission of microscopic influences. In contrast, most PDE discovery methods rely entirely on data-driven, nonparametric techniques and lack interpretable structures. \cite{yin2021augmenting} proposed an augmenting physical PDE model by incorporating a small perturbation treated as a penalty term to enhance representational capacity. 
These observations on the limitations of current PDE modeling approaches  motivate a new perspective on filling the need of combining the interpretability of physical mechanisms with the flexibility of data-driven methods to more accurately represent real-world scenarios.

In this paper, we propose Semiparametric Partial Differential Equations (SemiPDE) to model unknown mechanisms in data generation and evolution. The SemiPDE model consists of two components: a parametric part, which includes established physical laws that govern macroscopic data behavior, although certain physical parameters $\theta$ may remain unknown; and a nonparametric component $\mathcal{F}(u)$, which accounts for complex environmental factors or cognitive limitations not described by existing physical laws. The SemiPDE model integrates both known physical laws and data-driven techniques to capture complex generating mechanisms.
The nonparametric component is a multidimensional function that involves $x,u(x),\partial_x u(x)$ and possibly higher-order derivatives of $u(x)$. We consider estimating this  within a neural network space due to the strong approximation ability of neural networks for multidimensional functions. There are two main challenges in estimating and inferring the SemiPDE model. The first challenge is to develop a nonparametric estimator that achieves a sufficiently fast convergence rate while accounting for the structure of the SemiPDE model. 
The second challenge is how to perform inference on the unknown parameters $\theta$ without an explicit efficient score function due to no closed form for the solution $u(x)$ of the SemiPDE model, which also needs to mitigate the effect of the nonparametric function estimated by neural networks.


The main contributions of this work are summarized as follows. First, we propose the SemiPDE model, which effectively captures unknown mechanisms in data dynamics. The model is interpretable due to the inclusion of established physical laws, yet flexible  to accommodate trainable nonparametric components. Several examples are provided to illustrate the validity and general applicability of the SemiPDE model. Second, we develop a penalized profiling M-estimation procedure that decouples the solution $u(x)$ from the parameters in the SemiPDE model. Our approach incorporates a profiling strategy that leverages both the accuracy of numerical solutions for given PDEs and the expressive power of neural networks. This results in a more efficient and practical estimation method compared to Z-estimation or joint M-estimation techniques. Third, we derive optimal nonparametric convergence rates for the algorithmic estimators of both the unknown mechanisms and the solution $u(x)$, and establish root-$n$ consistency and asymptotic normality for the estimator of the (finite-dimensional) unknown parameter $\theta$ in physical laws.
This is the first systematic analysis of the statistical properties of neural network estimators that considers both training dynamics and the complex differential structure of PDE models. In the presence of neural networks, traditional semiparametric theory fails to achieve parametric efficiency without an explicit efficient score function. In contrast, our estimator for $\theta$ attains parametric efficiency for complex PDE models, overcoming the limitations of conventional semiparametric approaches. Finally, we develop a valid inference procedure for the parameter $\theta$ with theoretical guarantees. We propose a fast and easily implementable method to estimate the efficient covariance matrix, which transforms its intractable form into a manageable M-estimation problem. This inference procedure eliminates the need for complex differentiation or explicit construction of the score function. Numerical simulations and real data analyses further demonstrate the effectiveness of the proposed SemiPDE estimation and inference framework.

The rest of the paper is structured as follows. Section \ref{sec:model} introduces the proposed SemiPDE model, discusses the challenges in estimation and inference, and presents the penalized profiling M-estimators and algorithms for estimating the parametric and nonparametric components. Section \ref{sec:theorem_nonpara} establishes the convergence rates for the unknown mechanisms and the solutions to the equations. Section \ref{sec:theorem_para} analyzes the efficiency of the parameter estimators and proposes a valid inference procedure. Sections \ref{sec:simulation} and \ref{sec:realdata} demonstrate the strong performance of the SemiPDE model and its estimators in both simulation studies and real-data applications. Section \ref{sec:discussion} presents the conclusions of this paper and provides a brief discussion. Additional theoretical results, technical proofs and numerical analyses are provided in the online Supplementary Material due to space limitations.

\vspace{-1cm}

\section{Model and Methodology}
\label{sec:model}
\subsection{The SemiPDE model and examples}
We consider data-generating processes governed by underlying mechanisms, which are common in many scientific fields.
The observed data $(X_{i},Y_{i})$ with $X_i\in \mathbb{R}^{d_x},Y_i\in \mathbb{R}^{d_y}$ are in the form of 
\begin{align}
\label{eq:regression}
    Y_{i} = u(X_{i}) + \varepsilon_{i}, \quad i = 1,2,\cdots,n,
\end{align}
where $u(\cdot)$ is an unknown function, and $\varepsilon_{i}$, $1\le i \le n$, are independent zero-mean sub-exponential noise with variance $\sigma^2 I_{d_y}$. 
The governing mechanisms are modeled as a system of PDEs that includes both established physical laws and unknown factors influencing the data evolution, with $u(\cdot)$ as the solution.
Formally, we propose the SemiPDE model, defined as follows:
\begin{equation}
	\label{eq:PDEdifinition}
 \begin{aligned}
     \mathcal{L}(u(x);\theta) + \mathcal{F}(u(x)) &= 0,\quad x\in D\\
	\mathcal{B}(u(x)) &= 0,\quad x \in \partial D,
 \end{aligned}
\end{equation}
where the governing equations consist of two components $\mathcal{L}$ and $\mathcal{F}$ within the domain $D$, and known boundary and initial conditions $\mathcal{B}(u)=0$. In the SemiPDE model \eqref{eq:PDEdifinition}, $\mathcal{L}$ represents the known physical mechanisms, described by a differential operator within a parametric family indexed by a finite-dimensional parameter $\theta \in \Theta \subset \mathbb{R}^p$. In contrast, $\mathcal{F}$ represents the totally unknown mechanisms, modeled as a nonparametric differential operator of $u(\cdot)$.

Our SemiPDE model \eqref{eq:PDEdifinition} includes a range of realistic problems. Two representative examples are presented below.
\begin{emp}
    \label{emp:r-d}
    (Reaction-diffusion system)
    Let $\mathcal{L}$ represent the evolution and diffusion mechanisms $\mathcal{L}(u(t,x);\theta) = \partial_t u(t,x)  - \theta \Delta u(t,x)$, where $\partial_t u(t,x)$ denotes the temporal evolution and $\Delta u(t,x) = \partial_x^2 u(t,x)$ is the Laplacian operator representing diffusion. The parameter $\theta$ controls the diffusion rate. Let $\mathcal{F}$ denote the unknown local reaction mechanism given by $\mathcal{F}(u(t,x)) = f(u(t,x),t,x)$, where $f$ is an unknown function of $u(t,x),t,x$. Then \eqref{eq:PDEdifinition} models the reaction-diffusion system
    \begin{equation}
    \label{eq:R-D system}
    \begin{aligned}
        \partial_t u(t,x) - \theta\Delta u(t,x) + f(u(t,x),t,x)& = 0,\quad (t,x)\in D \\
        \mathcal{B}(u(t,x)) &=  0,\quad (t,x)\in \partial D.
    \end{aligned}
    \end{equation}
\end{emp}

\begin{emp}
    \label{emp: n-s}
   (Incompressible Navier–Stokes equations)  
   Let $\mathcal{L}$ represent the mechanisms of evolution, convection and diffusion defined as $\mathcal{L}(u(t,x);\nu,\rho) = \rho(\partial_t u(t,x) + u(t,x) \cdot \nabla u(t,x) - \nu \Delta u(t,x))$. {In this formulation, $\partial_t u(t,x)$ and $\Delta u(t,x) $ still denote the temporal evolution and diffusion mechanisms. The term $u(t,x) \cdot \nabla u(t,x)$ where $\nabla u(t,x)$ is the gradient of $u(t,x)$ with respect to $x$, represents convection. The parameters $\rho$ denotes density and $\nu$ denotes viscosity.} Let $\mathcal{F}$ denote the unknown source term $\mathcal{F}(u(t,x)) = f(t,x)$ which can represent the force term or the influence of pressure and stress. Then \eqref{eq:PDEdifinition} models the incompressible Navier–Stokes equations
   \begin{equation}
   \label{eq:incomNS}
       \begin{aligned}
           \rho(\partial_t u(t,x) + u(t,x) \cdot \nabla u(t,x) - \nu \Delta u(t,x)) + f(t,x) &= 0,\quad (t,x) \in D\\
           \mathcal{B}(u(t,x)) &= 0,\quad (t,x)\in \partial D,
       \end{aligned}
   \end{equation}
   which describe the motion of incompressible fluids. When $\mathcal{F}(u(t,x))=F(x)$, \eqref{eq:incomNS} reduces to the Burger's equation with an external force $F(x)$. 
\end{emp}

Unlike parametric models which assume that the mechanisms governing data dynamics are fully known (i.e., $\mathcal{F}=0$), our SemiPDE model provides greater flexibility. It captures complex influences on data dynamics through the data-driven term $\mathcal{F}$. In practice, the mechanisms underlying data generation and evolution are often not fully understood in advance. Including the nonparametric term $\mathcal{F}$ enhances model flexibility and may help the discovery of unknown mechanisms. On the other hand, our SemiPDE model differs from purely data-driven models that ignore existing mechanisms (i.e., $\mathcal{L}=0$). The latter must estimate all dynamics from noisy data and cannot take advantage of established physical laws. By integrating known and stable physical principles, our model reduces variability in estimating known components and yields more interpretable results.

\subsection{Challenges in estimation and inference}
The SemiPDE model is a semiparametric model that involves a parameter $\theta$ and a nonparametric nuisance parameter $\mathcal{F}$. Before presenting estimation algorithms, we point out the challenges arising in estimating and making inferences about parameter $\theta$. Because of the unique structure of the SemiPDE model, the traditional semiparametric methods such as Z-/M-estimators might fail.

For a Z-estimator, the parameter $\theta$ is typically estimated by $\tilde{\theta}$, which satisfies the equation
\begin{align*}
    \frac{1}{n}\sum_{i=1}^n \psi(X_i,Y_i;\tilde{\theta},\tilde{\mathcal{F}}) = 0,
\end{align*}
where $\psi$ is a known score function and $\tilde{\mathcal{F}}$ is an estimator of $\mathcal{F}$. As demonstrated in \cite{chernozhukov2018double}, constructing an orthogonal score function $\psi$ which has a zero Gateaux derivative with respect to $\mathcal{F}$ is often required to achieve $\sqrt{n}$-consistency and asymptotic normality of $\tilde{\theta}$. However, this is particularly difficult in the SemiPDE model due to lack of a closed-form solution $u(x;\theta,\mathcal{F})$ and the absence of an explicit relationship between $u(x;\theta,\mathcal{F})$ and parameters $(\theta, \mathcal{F})$. Moreover, the efficiency of $\tilde{\theta}$ is not guaranteed, as deriving the efficient score function faces similar challenges.

We next consider the joint M-estimator that seems a naive choice. In the SemiPDE model \eqref{eq:PDEdifinition}, $\mathcal{F}$ is a differential operator that that depends on $x,u,\partial_x u,\partial^2_x u$, and potentially higher-order derivatives of $u$. This nonparametric function is often multi-dimensional and we approximate it using feedforward neural network (FNN) due to its strong performance in nonparametric modeling. Recently, deep learning based PDE solvers such as Physics-informed Neural Networks \citep[PINNs,][]{raissi2019physics} have attracted considerable attention. With PINNs, the solution $u$ to \eqref{eq:PDEdifinition} can be estimated jointly with $\theta$ and $\mathcal{F}$ during data fitting by minimizing 
\begin{align}
\label{eq:pinn}
    \frac{1}{n}\sum_{i=1}^n \left\|Y_{i} - u(X_{i})\right\|_2^2 + \lambda_1 \int_{D} \|\mathcal{L}(u(x);\theta) + \mathcal{F}(u(x))\|_2^2 \ dx + \lambda_2 \int_{\partial D} \|\mathcal{B}(u(x))\|_2^2 \ dx. 
\end{align}
However, jointly estimating $u$ with $\theta, \mathcal{F}$ using this M-estimation framework often leads to slow convergence. This is due to the amplification of statistical errors in estimating $u$, as the differential operators $\mathcal{L}$  and $\mathcal{F}$ reduce the smoothness of $u$.

\subsection{Methodology and algorithm}
In the SemiPDE model, constructing a score function for Z-estimation is challenging due to the implicit form of the solution $u(x)$, while joint M-estimation is inefficient due to the model’s differential structure. To address these limitations, we propose a new estimation framework for the parameter $\theta$ and the nonparametric operator $\mathcal{F}$ using penalized M-estimation with a profiling strategy:
\begin{align}
\label{eq:M-estimator}
    \hat{\theta},\hat{\mathcal{F}} = \underset{\substack{\theta\in \Theta\\ \mathcal{F}(u)\in \mathcal{S}_{n}}}{\arg\min}  L_n(\theta,\mathcal{F}) =  \underset{\substack{\theta\in \Theta\\ \mathcal{F}(u)\in  \mathcal{S}_{n}}}{\arg\min} \frac{1}{n}\sum_{i=1}^n \left\|Y_{i} - \tilde{u}(X_{i};\theta,\mathcal{F})\right\|_2^2 + \lambda_n J(\mathcal{F}),
\end{align}
where $\tilde{u}(x;\theta,\mathcal{F})$ is the numerical solution to the SemiPDE model \eqref{eq:PDEdifinition} and $J(\mathcal{F})$ is a penalty term with tuning parameter $\lambda_n \ge 0$. The estimator for the function $u(x)$ is given by $\hat{u}(x) = \tilde{u}(x;\hat{\theta},\hat{\mathcal{F}})$. 

In \eqref{eq:M-estimator}, we focus on estimating parameters $\theta$ and $\mathcal{F}$ that influence the loss function $L_n(\theta,\mathcal{F})$. The unknown solution $u(x)$ is decoupled from parameters $\theta$ and $\mathcal{F}$ using the profiling strategy. Specifically, $u(x)$ is represented as the function $\tilde{u}(x;\theta,\mathcal{F})$ parameterized by $\theta$ and $\mathcal{F}$, which best satisfies the SemiPDE model for given values of these parameters. In contrast to full M-estimation methods such as \eqref{eq:pinn} which estimate $u(x)$ jointly with $\theta$ and $\mathcal{F}$, the separation of $u(x)$ utilizes the high accuracy of the numerical solution to a given PDE which is not affected by the statistical noise. As a result, our method reduces the degrees of freedom associated with estimating $u(x)$, thereby mitigating potential estimation bias and numerical instability that can arise from joint estimation. Moreover, it helps to confine the statistical error in estimating $\theta$ and $\mathcal{F}$ within the neural network space, limiting its propagation to $u(x)$, while still benefiting from the expressive capacity of neural networks. 

Denote the FNN space $\mathcal{S}_{L,W}$ with fixed depth $L$ and width $W = (m_0,m_1,\cdots,m_{L+1}), m_0 = d,m_{L+1} = d_y$ as
\begin{align}
\label{eq:fnn_definition}
    \mathcal{S}_{L,W} = \left\{\tilde{f}(\cdot;\phi) \Bigg| \ \tilde{f}(\cdot;\phi) = \mathcal{L}_{L} \circ  \relu  \circ \mathcal{L}_{L-1} \circ  \relu  \circ \mathcal{L}_{L-2} \circ  \relu  \circ \cdots \circ \mathcal{L}_{1} \circ  \relu  \circ \mathcal{L}_{0}(\cdot) \right\}, 
\end{align}
where $\mathcal{L}_{i}(\cdot)=\sqrt{\frac{l+1}{m_{i+1}}}(W_{i} \cdot+b_{i})$ is a multivariate linear function with $W_{i} \in \mathbb{R}^{m_{i+1} \times m_{i}}, b_{i} \in \mathbb{R}^{m_{i}}$, $x \in \mathbb{R}^{m_i}$. The nonlinear activation function $\relu$ is of the form $a \mapsto \max\{a, 0\}^l, l \in \mathbb{N}^{+}$, including both the rectified linear unit (ReLU) and the rectified power unit (RePU). For convenience, we define $s_{nn} = (d-1)/2 + l$, which can be intuitively interpreted as the smoothness of the neural network space. In \eqref{eq:fnn_definition}, $\phi$ is the vector consisting of all weights $W_i$ and biases $b_i$ in the neural network. We set $\mathcal{S}_{L,W}$ as an overparametrized FNN space where the dimensionality of weights $\phi$ exceeds the sample size $n$. This is a common practice in training neural networks nowadays due to its strong representational and learning capabilities. The function space $\mathcal{S}_n$ used for the estimation of $\mathcal{F}$ is constructed as
\begin{align*}
    \mathcal{S}_n = \mathcal{S}_{L,W(n)} - \tilde{f}(\cdot;\phi_0) = \left\{f(\cdot;\phi)\ \Bigg| \ f(\cdot;\phi) = \tilde{f}(\cdot;\phi) - \tilde{f}(\cdot;\phi_0)\right\},
\end{align*}
where $\tilde{f}(\cdot;\phi_0)\in\mathcal{S}_{L,W(n)}$ is a fixed function with predetermined weights $\phi_0$ consisting of $W_{0,i}\in \mathbb{R}^{m_{i+1}\times m_{i}},b_{0,i}\in \mathbb{R}^{m_i}$. One advantage of choosing $\mathcal{S}_n$ is that the optimization of \eqref{eq:M-estimator} can begin with $\mathcal{F}(u) =0$ when the weights are initialized at $\phi=\phi_0$. The penalty term $J(\mathcal{F})$ for $\mathcal{F}(u)=f(\cdot;\phi)\in \mathcal{S}_n$ is defined as
\begin{align*}
    J(\mathcal{F}) = J(f(\cdot;\phi)) = \|\phi - \phi_0\|_2^2.
\end{align*} 

\begin{figure}[htbp!]
\vspace{-0.5cm}
    \centering
    \includegraphics[width=1\linewidth]{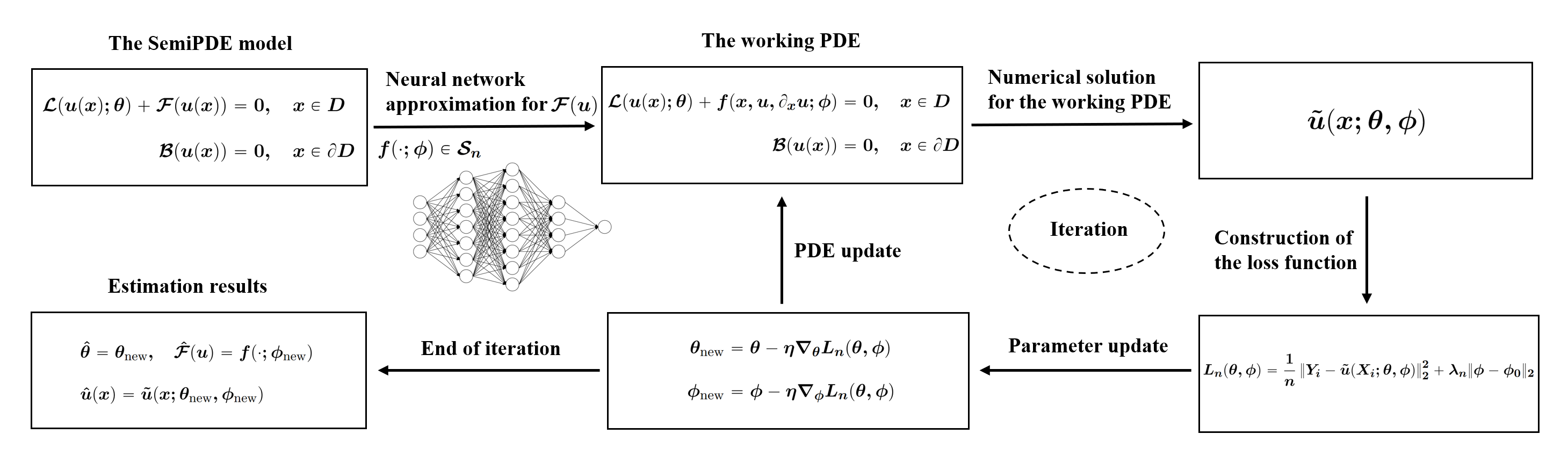}
    \caption{The estimation procedure of the SemiPDE model. }
    \label{fig:algorithm}
    \vspace{-1cm}
\end{figure}

A standard optimization approach is to identify $\mathcal{F}(u;\theta)\in\mathcal{S}_n$ that minimizes $L_n(\theta,\mathcal{F})$ for each $\theta$, and then determine the minimizer $\hat{\theta}$ of $\theta$. However, this is infeasible because a closed-form expression for $\tilde{u}(\cdot;\theta,\mathcal{F})$ is not available. Instead, we optimize \eqref{eq:M-estimator} using gradient descent, updating the parameter $\theta$ and the weights $\phi$ in the FNN jointly with the chain rule. The deviation of the solution $u(\cdot;\theta,\mathcal{F})$ with respect to $\theta$ and $\mathcal{F}$ is calculated using the ad-joint method \citep{giles2000introduction}. The deviation from $\mathcal{F}$ to the weights $\phi$ is calculated using the built-in automatic differentiation in neural networks. 
In summary, our estimation procedure is presented in Figure \ref{fig:algorithm} and Algorithm \ref{algo1}. Further details on optimization and implementation, including the selection of hyper-parameters such as depth $L$, width $W$, learning rate $\eta$, and tuning parameter $\lambda_n$, are provided in Sections S.1 and S.4 of the Supplementary Material.

\begin{algorithm}[htbp!]
\caption{SemiPDE Estimation}
\label{algo1}
\begin{algorithmic}[1]
\Require SemiPDE model \eqref{eq:PDEdifinition}, train data $(X_i,Y_i),i=1,\cdots,n$, test data $(X^{test}_i,Y^{test}_i),i=1,\cdots,n_{test}$ and maximum epochs $K$.
\Ensure Estimators $\hat{\theta}$ and $\hat{f}$ for unknown parameters and mechanisms.
\State For $k = 0$, choose initial parameters $\theta^{(0)} = \theta_0$, weights $\phi^{(0)} = \phi_0$ and Loss\_min = Inf;
\While{$k\le K$}
\State $\nabla_\theta L^{(k+1)} \leftarrow  \frac{2}{n}\sum_{i=1}^n \left(\tilde{u}(X_i;\theta^{(k)},f^{(k)}) - Y_i\right)^\top\partial_\theta u(X_i;\theta^{(k)},f^{(k)})$;
\State $\nabla_\phi L^{(k+1)} \leftarrow \frac{2}{n}\sum_{i=1}^n \left(\tilde{u}(X_i;\theta^{(k)},f^{(k)}) - Y_i\right)^\top\partial_\phi u(X_i;\theta^{(k)},f^{(k)}) +2\lambda_n \left(\phi^{(k)} -\phi_0\right)$;
\State $\theta^{(k+1)} = \theta^{(k)} - \eta \nabla_\theta L^{(k+1)}, \ \phi^{(k+1)} = \phi^{(k)} - \eta \nabla_\phi L^{(k+1)}$;
\State Loss\_current = $\frac{1}{n_{test}}\sum_{i=1}^{n_{test}} \left\|Y_i^{test} - \tilde{u}(X_i^{test};\theta^{(k+1)},f^{(k+1)})\right\|^2 +\lambda_n \|\phi^{(k+1)} - \phi_0\|_2^2$;
\If{Loss\_current $<$ Loss\_min} 
\State $\hat{\theta} = \theta^{(k+1)}, \hat{\mathcal{F}} = f(\cdot;\phi^{(k+1)})$;
\State Loss\_min = Loss\_current;
\EndIf
\EndWhile
\State \textbf{output} $\hat{\theta},\hat{\mathcal{F}}$.
\end{algorithmic}  
\end{algorithm}

\section{Optimal Nonparametric Convergence}
\label{sec:theorem_nonpara}
\subsection{Model assumptions}
Let the true governing model \eqref{eq:PDEdifinition} for the data dynamics have the parameter $\theta_0 \in \mathbb{R}^{p}$ and the differential operator $\mathcal{F}_0$. The true operator $\mathcal{F}_0$ can be represented by a function $f_0$, which maps a set of variables from $x_i,u_i,\partial_{x_j} u_i,\cdots,$ and other covariates $z$ to a compact subset in $\mathbb{R}^{d_y}$. We denote this set of variables as $\mathcal{V}$ which contains $d$ elements, such that $\mathcal{F}_0(u) = f_0(\mathcal{V})$. Let $u_0(x)$ represent the solution to the true SemiPDE model. 

First, we state the assumptions on the true parameter and nonparametric function in the SemiPDE model, as well as the neural network space $\mathcal{S}_n$ used for estimation.
\begin{assump}
        \label{assum:nonparametric_class}
The true parameter $\theta_0$ belongs to a tight subset $\Theta\subset \mathbb{R}^p$. The true function $f_0$  belongs to a Sobolev space $\mathcal{W}^{s,2}$ for some $s > 0$. 
\end{assump}

\begin{assump}
\label{assum:fnn_space}
Neural network structures in $\mathcal{S}_n$ and initial strategies for $\phi_0$ satisfies that
\begin{enumerate}
        \item[a.] The depth $L>0$ is fixed, and for the width $W$, the condition $m \le \min_{1\le i\le L} m_i \le \max_{1\le i \le L} m_i \le Cm$ holds, where $C\ge 1$ is a constant.
        \item[b.] For the predetermined weights $\phi_0$, the elements of $W_{0,i}$ and $b_{0,0}$ are drawn from i.i.d. standard normal distributions and all other biases $b_{0,i},i\ge 1$ are set to zero. 
    \end{enumerate}
\end{assump}

In classical nonparametric and semiparametric models, the Lipschitz and margin conditions are fundamental for linking perturbations in unknown parameters and functions to changes in the expected loss function. Similar conditions appear in PDE models, where the margin condition is often referred to as the stability condition. {Specifically, the Lipschitz condition ensures that the solution varies smoothly when the form of the PDE change. In contrast, the stability condition indicate whether, near the true PDE form, changes in the form lead to substantial changes in the solution. This reflects the local stability needed to estimate the PDE form from noisy data. Additionally, the sensitivity condition is often applied in PDE settings to ensure that parameters have a significant effect on the solution, which also relates to the sensitive analysis for PDEs.} We impose a general assumption that summarizes these conditions in our SemiPDE model \eqref{eq:PDEdifinition}.
\begin{assump}
    \label{assum:ture_solution}
    The solution $u(x;\theta,\mathcal{F})$ of the SemiPDE model \eqref{eq:PDEdifinition} satisfies that
    \begin{enumerate}
     \item[a.] (Lipschitz condition) For every $\theta_1,\theta_2\in \Theta$ and $\mathcal{F}_1(u)=f_1(\mathcal{V}),\mathcal{F}_2(u)=f_2(\mathcal{V})\in\mathcal{W}^{s_{nn},2}$, it holds for some constant $C_1\ge 0$ that
      \begin{align*}
        \|u(x;\theta_1,\mathcal{F}_1) - u(x;\theta_2,\mathcal{F}_2)\|_{L_2} &\le C_{1} \left(\|\theta_1 - \theta_2\|_2 + \|f_1(\mathcal{V}) - f_2(\mathcal{V})\|_{L_2}\right).
        \end{align*}
        This condition also holds when replacing $u$ to $\partial_\theta u,\partial_\mathcal{F} u$ or $\mathcal{V}$. 
        For constants $r, C_2\ge 0$ and some $r$th order integral operator $\Gamma^{r,\alpha}$ (the inverse operator of a $r$th differential operator $D^{\alpha} f(x_1,\cdots,x_l)= \partial^{\alpha_1}_{x_1}\partial^{\alpha_l}_{x_l} f(x_1,\cdots,x_l)$ with $\sum_{i=1}^l \alpha_i = r$) that
        \begin{align}
        \label{eq:Lipschitz_for_integral}
            \|u(x;\theta_1,\mathcal{F}_1) - u(x;\theta_2,\mathcal{F}_2)\|_{L_2} \le C_{2} \left(\|\theta_1 - \theta_2\|_2 + \|\Gamma^{r,\alpha}(f_1(\mathcal{V}) - f_2(\mathcal{V}))\|_{L_2}\right).
        \end{align}
         \item[b.] (Stability condition) For every $\theta\in \Theta$ and $\mathcal{F}(u)=f(\mathcal{V})\in\mathcal{W}^{s_{nn},2}$, it holds for any $r'$th order integral operator $\Gamma^{r',\alpha}$ with constant $r'\ge r$ and $C_{3}\ge 0$ that,
        \begin{align}
        \label{eq:Stability_condition}
            \|u(x;\theta,\mathcal{F}) - u(x;\theta_0,\mathcal{F}_0)\|_{L_2}^2 \ge C_3  \left(\|\theta - \theta_0\|^2_2 +\|\Gamma^{r',\alpha}(f(\mathcal{V})-f_0(\mathcal{V}))\|_{L_2}^2\right).
        \end{align}
         \item[c.] (Sensitivity condition) For every $\mathcal{F}(u)=f(\mathcal{V})\in \mathcal{W}^{s_{nn},2}$, it holds for some constant $C_\theta >0$ that 
         \begin{align*}
             \inf_{\theta \in \Theta}\|\partial_{\theta} u(x;\theta,\mathcal{F})\|_{L_2} \ge C_\theta.
         \end{align*}
    \end{enumerate}
\end{assump}

The Lipschitz condition \eqref{eq:Lipschitz_for_integral} and the stability condition \eqref{eq:Stability_condition} in Assumption \ref{assum:ture_solution} with $r=r'=0$ coincide with the classical Lipschitz condition and margin condition used in traditional nonparametric and semiparametric regression problems. However, for models involving derivatives and integrals, the conditions with $r=r'=0$ do not always hold. 
Studying these conditions is crucial in the context of forward and inverse problems in PDEs. In previous works on parametric PDEs, similar conditions are commonly assumed (e.g. Assumption 4.1 and Equation 5.28 in \cite{de2024numerical}) , and several widely used parametric PDE models have been shown to satisfy them. However, due to the inclusion of the data-driven nonparametric term, verifying these conditions for specific SemiPDE models remains an open and challenging problem. While this verification can be particularly difficult for complex systems, such as the three-dimensional Navier-Stokes equations, many standard PDEs exhibit stability in both forward and inverse problems. We briefly demonstrate that these conditions also apply to the SemiPDE model using a few illustrative examples. For instance, the reaction-diffusion equations \eqref{eq:R-D system} in Example \ref{emp:r-d} satisfy the required conditions, as shown in \cite{sakthivel2011inverse} and \cite{kaltenbacher2020inverse} under mild constraints on the unknown reaction mechanism. We also provide a concrete example of the SemiPDE model as follows, involving a two-dimension linear differential equation. 

\begin{emp}
       \label{emp:firstorder_linearequation}
    For the SemiPDE model,
    \begin{align}
    \label{eq:emp:firstorder_linearequation}
        \partial_x u(x) - \theta \cdot u(x) - \mathcal{F}(u) = 0, \quad x\in[0,1],\quad u(0) = 0,
    \end{align}
    with $\theta = (\beta,2\beta)^\top, \beta \in [0,2], \theta_0=(1,2)^\top$ and $\mathcal{F}(u) = (e^{x}f(x),e^{x}f(x))^\top, 2\ge f(x)\ge 1$, the Lipschitz condition \eqref{eq:Lipschitz_for_integral} for $r=1$ and the stability condition for $r'=1$ are hold.  
\end{emp}
The condition $r = r'$ is commonly satisfied in many SemiPDE models such as \eqref{eq:R-D system} and \eqref{eq:emp:firstorder_linearequation}. To improve the clarity and intuition of our results, we focus on the case where $r = r'$ in the sequel. Results for more general cases $r'\ge r$ are presented in Section S.6. of the Supplementary Material.

Lastly, we assume that the numerical solutions $\tilde{u}(X;\theta,\mathcal{F})$ are sufficiently accurate.
\begin{assump}
    \label{assum:numerical_error}
    For every $\theta\in \Theta$ and $\mathcal{F}(u)=f(\mathcal{V})\in\mathcal{W}^{s_{nn},2}$, the numerical solution $\tilde{u}(x;\theta,\mathcal{F})$ satisfies that $\|\tilde{u}(x;\theta,\mathcal{F}) - u(x;\theta,\mathcal{F})\|_{L_2} \le \varepsilon_u$ and 
        \begin{align*}
             \|\partial_\theta \tilde{u}(x;\theta,\mathcal{F}) - \partial_\theta u(x;\theta,\mathcal{F})\|_{L_2}+ \|\partial_\mathcal{F}\tilde{u}(x;\theta,\mathcal{F}) - \partial_\mathcal{F} u(x;\theta,\mathcal{F})\|_{L_2}\le \varepsilon_u,
        \end{align*}
        for a predetermined numerical error tolerance $\varepsilon_u$.
\end{assump}
It is important to note that this assumption concerns only the accuracy of the numerical solution of a given PDE, rather than any statistical perspective. Several numerical methods, such as finite difference methods and finite element methods, can provide the required numerical solutions.

\subsection{Nonparametric convergence rates}
Unlike conventional statistical analyses that focus solely on the properties of the ideal global minimum of the M-estimation \eqref{eq:M-estimator}, we study the actual training process and analyze the properties of the solutions generated by optimization algorithms. Without loss of generality, we analyze the gradient flow as a continuous approximation of the discrete gradient descent algorithm. Since the learning rate $\eta$ in Algorithm \ref{algo1} is sufficiently small at each iteration, the discrepancy between the results of gradient descent and gradient flow is minimal. Consider the training process starting from the initial values $\hat{\theta}_0$ and $\hat{\phi}_0 = \phi_0$ as
\begin{align*}
    \frac{d}{dt} \hat{\theta}_t &= -\eta \nabla_\theta L_n(\hat{\theta}_t,\hat{\mathcal{F}}_t) = -\frac{\eta}{n}\sum_{i=1}^n \left(\tilde{u}(X_i;\hat{\theta}_t,\hat{\mathcal{F}}_t) - Y_i\right)^\top \partial_\theta \tilde{u}(X_i;\hat{\theta}_t,\hat{\mathcal{F}}_t),\\
    \frac{d}{dt}\hat{\phi}_t &= -\eta \nabla_\phi L_n(\hat{\theta}_t,\hat{\mathcal{F}}_t) = -\frac{\eta}{n}\sum_{i=1}^n \left(\tilde{u}(X_i;\hat{\theta}_t,\hat{\mathcal{F}}_t) - Y_i\right)^\top \partial_{\mathcal{F}} \tilde{u}(X_i;\hat{\theta}_t,\hat{\mathcal{F}}_t) \partial_\phi \hat{\mathcal{F}}_t(u_{i,t}) - 2\eta\lambda_n(\hat{\phi}_t - \phi_0).
\end{align*}
Denote that $\hat{\mathcal{F}}_t(u) = \hat{f}_t(\mathcal{V}) = f(\mathcal{V};\hat{\phi}_t)$ and $\hat{\theta}_{\infty} = \lim_{t\rightarrow\infty}\hat{\theta}_t,\hat{\mathcal{F}}_{\infty} = \lim_{t\rightarrow\infty}\hat{\mathcal{F}}_t$.  The estimators $\hat{\theta}$ and $\hat{\mathcal{F}}$ can then be obtained by $\hat{\theta} = \hat{\theta}_T, \hat{\mathcal{F}} = \hat{\mathcal{F}}_T$, where the stopping time $T$ is chosen as 
\begin{align*}
    T = \inf_{t>0} \left\{\left|\frac{d}{dt} \left(L_n(\hat{\theta}_t,\hat{\mathcal{F}}_t) - L_n(\hat{\theta}_\infty,\hat{\mathcal{F}}_\infty)\right) \right| \le \tau\right\},
\end{align*}
for some tolerance $\tau>0$. We also denote that $\hat{f}(\mathcal{V}) = \hat{f}_T(\mathcal{V})$ and $\hat{\phi} = \hat{\phi}_T$. 

First, we present the upper bound on the error for the estimation in \eqref{eq:M-estimator} of the unknown mechanisms $\mathcal{F}_0=f_0(\mathcal{V})$. 

\begin{thm}
\label{thm:nonparametric_rate}
     Let $\tau/\eta \le n^{-2(s+r)/(2(s+r)+d)}, \lambda_n = n^{-2(s_{nn}+r)/(2(s+r)+d)}\log^{(s_{nn}+r)/(s+r)} n$ and the conditions of Assumptions \ref{assum:nonparametric_class}-\ref{assum:numerical_error} hold with $\varepsilon_u \le (\tau/\eta)^{C}, m\ge (\eta/\tau)^{6C}$ for a constant $C>0$. If $s_{nn} \ge s$, the estimator $\hat{\mathcal{F}}(u) = \hat{f}(\mathcal{V})$ satisfies that
    \begin{align*}
        \|\hat{f}(\mathcal{V}) - f_0(\mathcal{V})\|^2_{L_2} = O_p\left( n^{-2 s/(2(s+r)+d)}\log n\right).
    \end{align*}
\end{thm}

Theorem \ref{thm:nonparametric_rate} establishes the first statistical properties of algorithmic solutions for neural network estimation of nonparametric functions in complex PDE models. When the Lipschitz condition \eqref{eq:Lipschitz_for_integral} and the stability condition \eqref{eq:Stability_condition} in Assumption \ref{assum:ture_solution} hold with $r=0$,
the convergence rate for $\hat{f}$ matches the optimal nonparametric rate $n^{-2s/(2s+d)}$ \citep{stone1980optimal}, up to a logarithm factor $\log n$ introduced by the use of neural networks \citep{schmidt2020nonparametric}. The result in Theorem \ref{thm:nonparametric_rate} also matches the minimax rate for nonparametric derivative estimation. In the derivative estimation problem, the observations $(X_i,Y_i),1\le i\le n$ follow $Y_i = u(X_i)+\varepsilon_i$, in which $f(x) = u^{(r)}(x)$ is the $r$th derivative of $u(x)$. For this problem, the Lipschitz condition and the stability condition in Assumption \ref{assum:ture_solution} hold. Thus our result in Theorem \ref{thm:nonparametric_rate} is consistent with the optimal convergence rate for derivative estimation, which is $n^{-2s/(2(s+r)+d)}$ \citep{de2013derivative,dai2016optimal}.

The convergence rate for $\hat{f}(\mathcal{V})$ is not effected by the choice of $\mathcal{S}_n$. This indicates that our method adapts to the smoothness of true function $f_0(\mathcal{V})$ as long as $s_{nn}\ge s$ holds. If the smoothness of the space $\mathcal{S}_n$ is less than $f_0$, i.e., $s_{nn}< s$, the results in Theorem \ref{thm:nonparametric_rate} no longer apply. In this case, the convergence rate is given by $n^{-2s_{nn}/(2(s_{nn}+r)+d)}\log n$, which is limited by the lower smoothness of the space  $\mathcal{S}_n$. In Theorem \ref{thm:nonparametric_rate}, a larger $r$ leads to a slower rate for $\hat{f}$. This is due to the perturbation of $f$ having little impact on the change in $u(x;\theta,f)$ when $r$ is large. This typically occurs in the estimation of high-order derivative and high-order PDEs. 

Based on the estimators $\hat{\theta}, \hat{f}$ defined in \eqref{eq:M-estimator}, we estimate the solution $u_0(x)$ sequentially by $\hat{u}(x) = \tilde{u}(x;\hat{\theta},\hat{f})$. Theorem \ref{thm:solution_rate} presents the error bound for the estimated solution $\hat{u}(x)$.  
\begin{thm}
    \label{thm:solution_rate}
    Let $\tau/\eta \le n^{-2(s+r)/(2(s+r)+d)}, \lambda_n = n^{-2(s_{nn}+r)/(2(s+r)+d)}\log^{(s_{nn}+r)/(s+r)} n$ and the condition of Assumptions \ref{assum:nonparametric_class}-\ref{assum:numerical_error} hold with $\varepsilon_u \le (\tau/\eta)^{C}, m\ge (\eta/\tau)^{6C}$ for a constant $C>0$. If $s_{nn} \ge s$, the estimator $\hat{u}(x)$ for true solution $u_0(x)$ satisfies that
    \begin{align*}
        \|\hat{u}(x) - u_0(x)\|^2_{L_2} = O_p\left( n^{-2 (s+r)/(2(s+r)+ d)}\log n\right).
    \end{align*}
\end{thm}
In classical nonparametric regression problems with $r = 0$, the estimator $\hat{u}$ still achieves the optimal rate $n^{-2s/(2s+d)}$. For the derivative estimation problem, the minimax convergence rate for $\hat{u}$ is $n^{-2(s+r)/(2(s+r)+d)}$, which aligns with this result. In the SemiPDE model where $r>0$, the rate for $\hat{u}$ is faster than that for $\hat{f}$ as stated in Theorem \ref{thm:nonparametric_rate}. This discrepancy arises primarily because the estimation of $f_0$ is related to the derivative of $u_0$, which is slower than to estimate $u_0$ itself.

\section{Efficient Inference Procedure}
\label{sec:theorem_para}
\subsection{Parametric efficiency in the SemiPDE model}

To assess the properties of the estimator $\hat{\theta}$, we must evaluate the influence of the nuisance parameter $f$ on the solution $u(x;\theta,f)$ and ensure that the slower nonparametric rate of the estimator $\hat{f}$ does not significantly affect the faster parametric rate of $\hat{\theta}$. The influence can be controlled if there exists an orthogonal direction $h_0$ along which the nuisance parameter has limited impact. The following assumptions formally establish the existence of such a sufficiently smooth direction $h_0$. Compared to previous works such as \cite{chernozhukov2018double} which require the explicit form of $h_0$, our condition is weaker, as it only assumes the existence of $h_0$. 

\begin{assump}
    \label{ass:neyman_orthogonal}
There exists $h_0\in \mathcal{W}^{s',2}$ for some $s'> d/2$ such that, for any $h\in \mathcal{W}^{s,2}$,
\begin{align}
\label{eq:ass:neyman_orthogonal}
        \mathbb{E}_X \left[\left(\partial_\theta u(X;\theta_0,f_0) - \partial_f u(X;\theta_0,f_0)[h_0]\right) \partial_f u'(X;\theta_0,f_0)[h]\right] = 0.
\end{align}
\end{assump}

\begin{assump}
    \label{ass:efficient_orthogonal}
    Let $h_{0} = -\partial_\theta f(\mathcal{V};\theta_0)$, where
    \begin{align*}
        f(\mathcal{V};\theta) = \arg\min_{f\in \mathcal{W}^{s,2}} \mathbb{E}_{X} \left\|u(X;\theta,f) - u_0(X)\right\|_2^2.
    \end{align*}
    Then $h_0 \in \mathcal{W}^{s',2}$ for some $s'> d/2$. Additionally, the noises term $\varepsilon_i$ in \eqref{eq:regression} follow Gaussian distributions. 
\end{assump}

\vspace{-0.5cm}
Denote that $l(\theta,f;x,y) = \left\|y - u(x;\theta,f)\right\|_2^2$ and
    \begin{align*}
        \psi(\theta,f;x,y) = \partial_\theta l(\theta,f;x,y) - \partial_f l(\theta,f;x,y)[h_0],\quad \mathbb{P}\psi(\theta,f) = \mathbb{E}_{X,Y} \psi(\theta,f;X,Y). 
    \end{align*}
The specific $h_0$ in Assumption \ref{ass:efficient_orthogonal} ensures that Assumption \ref{ass:neyman_orthogonal} holds with the same $h_0$. 
Both Assumptions \ref{ass:neyman_orthogonal} and \ref{ass:efficient_orthogonal} require a sufficiently smooth direction function $h_0$, which satisfies condition \eqref{eq:ass:neyman_orthogonal}. This condition \eqref{eq:ass:neyman_orthogonal} is essential for the score function $\psi(\theta_0,f_0;x,y)$ to have a zero Gateaux derivative with respect to the nonparametric nuisance parameter $f(\cdot)$. Unlike Z-estimations, our approach does not require explicitly constructing the score function $\psi$, which is infeasible for most SemiPDE models. Instead, we only need the existence of such an $h_0$ to ensure the existence of the orthogonal score function $\psi$. We provide an example to demonstrate the reasonableness of these assumptions for our SemiPDE model. Still consider the SemiPDE model \eqref{eq:emp:firstorder_linearequation} in Example \ref{emp:firstorder_linearequation}, it can be checked that $h_0(x) = \int_0^x f_0(s) ds \in \mathcal{W}^{s+1,2}$ satisfies Assumption \ref{ass:neyman_orthogonal}. 


Define the tangent space
$
\mathcal{T}_{\hat{\phi}} \mathcal{S}_n = \left\{g(\cdot)\ |\ g(\cdot) = \sum_i \gamma_i \partial_{\phi_i} f(\cdot;\hat{\phi})\right\}.
$
According to the first-order conditions, only the gradients
\begin{align*}
    \sum_{i=1}^n \partial_\theta l(\hat{\theta},\hat{f};X_i,Y_i),\quad \sum_{i=1}^n \partial_f l(\hat{\theta},\hat{f};X_i,Y_i)[h], h \in \mathcal{T}_{\hat{\phi}}\mathcal{S}_n
\end{align*}
 are guaranteed to be sufficiently small. However, we cannot ensure that $h_0$ lies in the tangent space $\mathcal{T}_{\hat{\phi}}\mathcal{S}_n$ and thus we cannot directly verify that the empirical average of the orthogonal score $\psi(\hat{\theta},\hat{f};x,y)$ is small. 
The tangent space $\mathcal{T}_\phi \mathcal{S}_n$ for a FNN $f(\cdot;\phi)\in \mathcal{S}_n$ is complex due to the nonlinearity of $\mathcal{S}_n$, making it challenging to analyze its approximation properties. In \cite{yan2025semiparametric}, this issue is also identified as a general problem encountered in semiparametric M-estimation using neural networks. Adopting similar techniques in \cite{yan2025semiparametric} to show that $h_0$ can be sufficiently approximated by functions in $\mathcal{T}_{\hat{\phi}} \mathcal{S}_n$, we need to address additional challenges caused by lack of a closed-form solution and the complex differential structures of the SemiPDE model.
 The next theorem establishes the root-$n$ consistency of $\hat{\theta}$ and its asymptotic normality. 

\begin{thm}
    \label{thm:parametric_rate}
    Let $\tau/\eta \le n^{-((2r+d)s_{nn}+2ss')/(2(s+r)+d)s'},\lambda_n = n^{-2(s_{nn}+r)/(2(s+r)+d)}\log^{(s_{nn}+r)/(s+r)} n$ and the condition of Assumptions \ref{assum:nonparametric_class}-\ref{ass:neyman_orthogonal} holds with $\varepsilon_u \le (\tau/\eta)^{C}, m\ge (\eta/\tau)^{6C}$ for a constant $C>0$. If the smallest singular value of $\partial_\theta\mathbb{P}\psi(\theta_0,f_0)$ is larger than a constant $c>0$ and the smoothness of $f_0$ and $\mathcal{S}_n$ satisfies $ s \wedge s_{nn} > r+ d/2$, the estimator $\hat{\theta}$ satisfies that 
    \begin{align*}
        n^{1/2}\left(\hat{\theta} - \theta_0\right) = - n^{-1/2}  \left(\partial_\theta^\top\mathbb{P}\psi(\theta_0,f_0)\right)^{-1}\sum_{i=1}^n \psi(\theta_0,f_0;X_i,Y_i) + o_p(1) \overset{d}\rightarrow N(0,\Sigma), 
    \end{align*}
    where
    \begin{align*}
        \Sigma = \left(\partial_{\theta^\top} \mathbb{P}\psi(\theta_0,f_0)\right)^{-1}\mathbb{E}_{X,Y}\left[\psi(\theta_0,f_0;X,Y)\psi(\theta_0,f_0;X,Y)^\top\right]\left(\partial_{\theta^\top} \mathbb{P}\psi(\theta_0,f_0)\right)^{-1}.
    \end{align*}
    Besides, if Assumption \ref{ass:efficient_orthogonal} holds instead of Assumption \ref{ass:neyman_orthogonal}, $\psi(\theta,f;x,y)$ is the efficient score function and the estimator $\hat{\theta}$ is efficient with 
    \begin{align*}
        n^{1/2}\left(\hat{\theta}-\theta_0\right)\overset{d}{\rightarrow} N(0,\Sigma_{\operatorname{eff}}).\,\quad \Sigma_{\operatorname{eff}} = \sigma^4\left(\mathbb{E}_{X,Y} [(\partial_{\theta}l(\theta_0,f(\cdot;\theta_0);X,Y)) (\partial_{\theta}l(\theta_0,f(\cdot;\theta_0);X,Y)^\top]\right)^{-1}.
    \end{align*}
\end{thm}

In Theorem \ref{thm:parametric_rate}, the invertibility of $\partial_\theta \mathbb{P}\psi(\theta_0,f_0)$ is assumed to prevent a degenerate scenario, which is a common assumption in semiparametric problems. If Assumption \ref{ass:efficient_orthogonal} is not satisfied, $\psi(\theta,f;x,y)$ may not be the efficient score function. But the estimator $\hat{\theta}$ remains root-$n$ consistent and asymptotically normal, with the variance $\Sigma$ slightly larger than the efficient variance $\Sigma_{\operatorname{eff}}$. If Assumption \ref{ass:efficient_orthogonal} holds, Theorem \ref{thm:parametric_rate} demonstrates that our estimation achieves parametric efficiency. To our knowledge, this is the first result that establishes the efficiency of the estimator of the parameter $\theta$ in semiparametric M-estimation for PDE models, where the function space is represented by FNNs with irregular tangent spaces.
\subsection{Inference procedure and its properties}

For inference, an additional estimation of the variance is required. The asymptotic variance $\Sigma$ or $\Sigma_{\operatorname{eff}}$ in Theorem \ref{thm:parametric_rate} is implicitly defined because of the unknown direction $h_0$ and score function $\psi$. Due to lack of a closed form of $u(x;\theta,f)$, identifying $h_0$ that satisfies Assumption \ref{ass:neyman_orthogonal} or \ref{ass:efficient_orthogonal} is challenging. Directly verifying \eqref{eq:ass:neyman_orthogonal} in Assumption \ref{ass:neyman_orthogonal} or estimating the gradient of the profiling function $f(\cdot;\theta)$ in Assumption \ref{ass:efficient_orthogonal} is unstable and computationally complex. 

{In contrast, we consider the following strongly convex optimizing problem 
\begin{align*}
    \min_{h \in \mathcal{W}^{s',2}} \mathbb{E}_{X} \left\| \partial_\theta u(X;\theta_0,f_0) - \partial_f u(X;\theta_0,f_0)[h]\right\|_2^{\otimes 2}.
\end{align*}
The condition \eqref{eq:ass:neyman_orthogonal} exactly implies that $h_0$ satisfies the first order condition of the above problem, indicating that $h_0$ is its minimizer. We can therefore reformulate $\Sigma_{\text{eff}}$ in the Theorem \ref{thm:parametric_rate} as 
\begin{align}
    \Sigma_{\text{eff}}& = \sigma^2 \left(\mathbb{E}_{X} \left\| \partial_\theta u(X;\theta_0,f_0) - \partial_f u(X;\theta_0,f_0)[h_0]\right\|_2^{\otimes 2}\right)^{-1} \notag\\&=
     \underset{h\in \mathcal{W}^{s',2}}{\arg\max}\ \sigma^2 \left(\mathbb{E}_{X} \left\| \partial_\theta u(X;\theta_0,f_0) - \partial_f u(X;\theta_0,f_0)[h]\right\|_2^{\otimes 2}\right)^{-1}. \label{eq:reform_variance}
\end{align}
Based on \eqref{eq:reform_variance}, we propose a stable and practical inference procedure. 
The dataset is divided into two parts: one is used to estimate $\theta_0$ and $f_0$ as $\hat{\theta}$ and $\hat{f}$ respectively, and the other is used to approximate the expectation in \eqref{eq:reform_variance}. The efficient variance $\Sigma_{\operatorname{eff}}$ is then estimated using finite differences as}
\begin{align}
    \label{eq:estimation_for_variance}
    \hat{\Sigma}_{\operatorname{eff}} = \delta^{2} \hat{\sigma}^2 \left(\frac{1}{n}\sum_{i=1}^n\left\|\tilde{u}(X_i;\hat{\theta}+\delta I_{p},\hat{f}) - \tilde{u}(X_i;\hat{\theta},\hat{f} + \hat{g})\right\|_{2}^{\otimes 2}\right)^{-1},
\end{align}
where $|\delta| =o(n^{-1/2})$, $\hat{\sigma}^2 =n^{-1}\sum_{i=1}^n \left\|Y_i - \tilde{u}(X_i;\hat{\theta},\hat{f})\right\|_2^2 $ is the estimation of $\sigma^2$ and 
\begin{align}
\label{eq:estimation_for_direction}
    \hat{g} = \underset{g\in \mathcal{S}_n^{\otimes p}}{\arg\min}\frac{1}{n} \sum_{i=1}^n \left\|\tilde{u}(X_i;\hat{\theta}+\delta I_p ,\hat{f}) - \tilde{u}(X_i;\hat{\theta},\hat{f} + \hat{g})\right\|_2^{\otimes 2} + \tilde{\lambda}_n J(g).
\end{align}
Our estimation of $\Sigma_{\operatorname{eff}}$ circumvents the direct estimation of the direction $h_0$ or the construction of the efficient score function, both of which are particularly challenging in the SemiPDE model. In practice, it is unnecessary to substitute the estimator $\hat{g}$ into \eqref{eq:estimation_for_variance}. Instead, only the value of the first term in \eqref{eq:estimation_for_direction} is required, which further reduces the computational complexity.

The next theorem demonstrates that the estimated variance $\hat{\Sigma}_{\operatorname{eff}}$ provides a simultaneous asymptotic $1-\alpha$ confidence interval for any linear combination $\gamma^\top\theta_0$ of $\theta_0$. 

\begin{thm}
    \label{thm:parametric_inference}
    Under the same conditions as in Theorem \ref{thm:parametric_rate} and assuming Assumption \ref{ass:efficient_orthogonal} holds, for any size $\alpha \in [0,1]$, we have the following result
    \begin{align*}
        P\left(\gamma^\top \hat{\theta} - n^{-1/2}z_{1-\alpha/2} \gamma^\top\hat{\Sigma}_{\operatorname{eff}}\gamma\le \gamma^\top\theta_0  \le \gamma^\top \hat{\theta} + n^{-1/2}z_{1-\alpha/2} \gamma^\top\hat{\Sigma}_{\operatorname{eff}}\gamma, \gamma \in \mathbb{S}^{p - 1}\right) = 1 - \alpha + o(n^{-1/4}),
    \end{align*}
    where $z_{1-\alpha/2}$ is the $1-\alpha/2$ quantile of the standard Gaussian distribution. 
\end{thm}
 Theorem \ref{thm:parametric_inference} provides a practically useful and minimal-length confidence interval for the true parameter $\theta_0$. The confidence interval $[\gamma^\top\hat{\theta} - n^{-1/2}z_{1-\alpha/2} \gamma^\top\hat{\Sigma}_{\operatorname{eff}}\gamma, \gamma^\top\hat{\theta} + n^{-1/2}z_{1-\alpha/2} \gamma^\top\hat{\Sigma}_{\operatorname{eff}}\gamma]$ achieves an asymptotic coverage probability of $1-\alpha$ for $\gamma^\top\theta_0$ with a discrepancy of order $o(n^{-1/4})$.
 
\section{Simulation}
\label{sec:simulation}
We apply the SemiPDE framework to several well-known problems to assess its effectiveness. Four special cases of governing equations which represent common dynamics in data generation and evolution are presented:
\begin{align*}
    \text{Case 1}:& \quad \mathcal{L}(u(t,x);\theta) = \theta \Delta u -  \partial_t u , \mathcal{F}(u(t,x)) = f(u) = u(1-u). \\
    \text{Case 2}:& \quad \mathcal{L}(\bm{u}(t,x);\bm{\theta}) = \left(\begin{matrix}
        \theta_1 \Delta u_1 -  \partial_t u_1 \\ \theta_2 \Delta u_2 - \partial_t u_2
    \end{matrix}\right),\mathcal{F}(\bm{u}(t,x)) = \left(\begin{matrix}
       f_1(\bm{u})\\f_2(\bm{u})
    \end{matrix}\right) =\left(\begin{matrix}
       u_1-u_1^3-5e^{-3}-u_2\\u_1-u_2
    \end{matrix}\right),\\
    \text{Case 3}:&\quad \mathcal{L}(u(t,\bm{x});\bm{\theta}) = \theta_1 \nabla u + \theta_2\nabla\cdot (\bm{E}u) -\partial_t u,\\
    & \quad \mathcal{F}(u(t,\bm{x})) = -\nabla \cdot \left(\begin{pmatrix}
        f_1(\bm{x})\\f_2(\bm{x})\\
    \end{pmatrix}u\right) = -\frac{1}{10}\nabla\cdot\left(\begin{pmatrix}
        (1-|x_1|)^2+(1-|x_2|)\\(1-|x_2|)^2+(1-|x_1|)\\
    \end{pmatrix} u\right),
        \end{align*}
    \begin{align*}
    \text{Case 4}:& \quad \mathcal{L}(\bm{u}(t,\bm{x});\bm{\theta}) = \partial_t \bm{u} + \bm{u}\nabla \bm{u} - \bm{\theta}_1\left(\Delta \bm{u}- \frac{1}{3}\nabla(\nabla\cdot \bm{u})\right)+ \bm{\theta}_2 \nabla\bm{p},\\&\quad\mathcal{F}(\bm{u}(t,\bm{x})) = -\bm{f}_1(\bm{x}) -\nabla f_2(\bm{x},\bm{u},\nabla \cdot \bm{u}),\\
    &\quad \bm{f}_1(\bm{x}) = \frac{1}{10}(\sin(2\pi(|x_1|+|x_2|+|x_3|)) +\cos(2\pi(|x_1| + |x_2| + |x_3|)))(1,1,1)^\top\\
    & \quad f_2(\bm{x},u,\nabla \cdot \bm{u}) = \frac{(3-|x_1|-|x_2|-|x_3|) \nabla \cdot \bm{u}}{10(1+u_1^2+u_2^2+u_3^2)}.
\end{align*}
Bold notation is used to distinguish vectors from scalars. The symbols $\Delta$, $\nabla$ and $\nabla\cdot$ denotes the Laplacian, gradient and divergence operators with respect to spatial variables, respectively. Case 1 and Case 2 are two reaction-diffusion equations, Case 3 is the Nernst-Planck equation where $\bm{E}$ represents the electric field strength and Case 4 is the 3D Navier-Stokes equation where $\bm{p}$ represents the pressure field. These models represent important governing equations, with detailed explanations and settings provided in Section S.2. of the Supplementary Material. The initial conditions are randomly generated as smoothness functions of $x$ or $\bm{x}$ in each case, with period boundary conditions applied. Data $(X_i,Y_i),1\le i\le n$, are generated by $Y_i = u(x_i;\theta,\mathcal{F})+\varepsilon_i$, where $X_i = (t_i,\bm{x}_i)$ is uniformly sampled from $[0,2.5]$ for the temporal variable $t$ and $[-1,1]$ for the spatial variables. The noise term $\varepsilon_i$ is independently and identically distributed according to a normal distribution with the standard deviation $\sigma$. The dataset is split into training and validation sets with a ratio of $4:1$. The tuning parameter $\lambda_n$ is selected based on the validation loss. For neural network construction, the activation function $\bm{\sigma}(\cdot)$ is set as ReLU: $a \mapsto \max\{a,0\}$ and four hidden layers with width $[16,64,64,16]$ are used. More implementation details are provided in Section S.4. of the Supplementary Material.

First, we demonstrate the performance of our SemiPDE method by comparing it with three baselines: Baseline 1 which uses parametric PDEs regression methods as in \cite{cao2012penalized} and \cite{xun2013parameter}; Baseline 2 which employs nonparametric regression similar to that in \cite{schmidt2020nonparametric}; Baseline 3 which integrates PINNs \citep{raissi2019physics} into the SemiPDE model. For Baseline 1, the known mechanism $\mathcal{L}(u;\theta)$ is used to construct the parametric PDE model $\mathcal{L}(u;\theta) =0$ and the parameter $\theta$ is estimated by 
\begin{align*}
    \tilde{\theta}_{1} = \underset{\theta\in \Theta}{\arg\min}\frac{1}{n}\sum_{i=1}^n \|Y_i - \tilde{u}(X_i;\theta)\|_2^2,
\end{align*}
where $\tilde{u}(x;\theta)$ is the numerical solution of the parametric PDE model. The solution $u(x)$ is then estimated as $\tilde{u}_1 = \tilde{u}(x;\tilde{\theta}_1)$. For Baseline 2, no mechanism is assumed and $u$ is estimated nonparametrically as
\begin{align*}
    \tilde{u}_2 = \underset{u\in \mathcal{S}_{L,W}}{\arg\min} \frac{1}{n}\sum_{i=1}^n \|Y_i - u(X_i)\|_2^2,
\end{align*}
where $\mathcal{S}_{L,W}$ denotes the FNN space \eqref{eq:fnn_definition} with proper depth $L$ and width $W$. For Baseline 3, the estimators for $\theta,\mathcal{F}$ and $u$ are jointly estimated by
\begin{align*}
    \tilde{\theta}_3,\tilde{\mathcal{F}}_3,\tilde{u}_3 = \underset{\substack{\theta\in \Theta, \mathcal{F}\in \mathcal{S}_n\\ u \in \mathcal{S}_{L,W}}}{\arg\min} \frac{1}{n}\sum_{i=1}^n \|Y_i - u(X_i)\|_2^2 + \lambda_1 \int_D \|\mathcal{L}(u(x);\theta) + \mathcal{F}(u(x))\|_2^2 \ dx. 
\end{align*}
Table \ref{tab:simu:compared_baseline} presents the estimation errors $\|\hat{u} - u_0\|_{L_2}$ for $u$ and $\|\hat{\theta} - \theta_0\|_2$ for $\theta$ from our SemiPDE method and three baselines. We consider scenarios with different sample sizes and noise levels in each case. The results are presented as the average of $50$ repeated experiments for each scenario. 
In Table \ref{tab:simu:compared_baseline}, our SemiPDE model demonstrates superior performance in estimating both parameter $\theta$ and the solution $u(x)$ compared to the baselines. By integrating the known mechanism $\mathcal{L}(u;\theta)$ with the unknown mechanism determined by observed data, our SemiPDE model is more suitable than both the mechanism-driven Baseline 1 and the data-driven Baseline 2. The proposed penalized M-estimators also outperform the joint estimation Baseline 3 in both mechanism estimation and solution approximation. 
\begin{table}[ht]
	\centerfloat
     \caption{Estimation errors for SemiPDE and Baselines.}
	\label{tab:simu:compared_baseline}
	\begin{tabular}{cc|cccc|cccc}
		\toprule[1pt]
		   Case & Method & \multicolumn{4}{c|}{Estimation error ($\times 10^{-2}$) for $u$}&  \multicolumn{4}{c}{Estimation error ($\times 10^{-2}$) for $\theta$} \\
     \midrule[1pt]
        \multicolumn{2}{c|}{\multirow{2}{*}{Settings}} &  \begin{footnotesize}
         $n$=800
     \end{footnotesize} & \begin{footnotesize}$n$=800\end{footnotesize}  &\begin{footnotesize} $n$=1600 \end{footnotesize} &\begin{footnotesize} $n$=1600 \end{footnotesize} &\begin{footnotesize}$n$=800 \end{footnotesize} & \begin{footnotesize}$n$=800 \end{footnotesize} &\begin{footnotesize} $n$=1600 \end{footnotesize} & \begin{footnotesize}$n$=1600\end{footnotesize}  \\
     & &\begin{footnotesize} $\sigma = 0.1$\end{footnotesize} & \begin{footnotesize}$\sigma = 0.5$ \end{footnotesize}&\begin{footnotesize} $\sigma = 0.1$\end{footnotesize}& \begin{footnotesize}$\sigma = 0.5$\end{footnotesize}&\begin{footnotesize}$\sigma = 0.1$ \end{footnotesize}&\begin{footnotesize} $\sigma = 0.5$ \end{footnotesize}&\begin{footnotesize} $\sigma = 0.1$\end{footnotesize}&\begin{footnotesize} $\sigma = 0.5$\end{footnotesize} \\
     \hline
     \multirow{4}{*}{ \begin{footnotesize}Case 1\end{footnotesize}} & \begin{footnotesize}SemiPDE\end{footnotesize} &  $\bm{0.261}$ & $\bm{0.971}$ & $\bm{0.192}$ & $\bm{0.766}$ & $\bm{0.032}$ & $\bm{0.116}$ & $\bm{0.031}$ & $\bm{0.084}$ \\
     &  \begin{footnotesize}Baseline 1 \end{footnotesize}&  20.86 & 20.86 & 20.48 & 20.47 & 1.108 & 1.148 & 0.974 & 1.112 \\
     & \begin{footnotesize}Baseline 2 \end{footnotesize}&  2.458  & 10.48 & 1.857 & 7.170 & \textbf{--} & \textbf{--} & \textbf{--}& \textbf{--} \\
     & \begin{footnotesize}Baseline 3 \end{footnotesize}&  2.348  & 10.73 & 2.626 & 10.60 &  1.314 & 3.203 &1.270& 3.896\\
     \hline
        \multicolumn{2}{c|}{\multirow{2}{*}{Settings}} &  \begin{footnotesize}
         $n$=3200
     \end{footnotesize} & \begin{footnotesize}$n$=3200\end{footnotesize}  &\begin{footnotesize} $n$=12800 \end{footnotesize} &\begin{footnotesize} $n$=12800 \end{footnotesize} &\begin{footnotesize}$n$=3200\end{footnotesize} & \begin{footnotesize}$n$=3200 \end{footnotesize} &\begin{footnotesize} $n$=12800 \end{footnotesize} & \begin{footnotesize}$n$=12800\end{footnotesize}  \\
     & &\begin{footnotesize} $\sigma = 0.1$\end{footnotesize} & \begin{footnotesize}$\sigma = 0.5$ \end{footnotesize}&\begin{footnotesize} $\sigma = 0.1$\end{footnotesize}& \begin{footnotesize}$\sigma = 0.5$\end{footnotesize}&\begin{footnotesize}$\sigma = 0.1$ \end{footnotesize}&\begin{footnotesize} $\sigma = 0.5$ \end{footnotesize}&\begin{footnotesize} $\sigma = 0.1$\end{footnotesize}&\begin{footnotesize} $\sigma = 0.5$\end{footnotesize} \\
     \hline
     \multirow{4}{*}{\begin{footnotesize}Case 2\end{footnotesize}} & \begin{footnotesize}SemiPDE \end{footnotesize}&  $\bm{0.397}$ &  $\bm{0.980}$ &  $\bm{0.331}$ & $\bm{0.588}$ & $\bm{0.022}$ & $\bm{0.068}$ & $\bm{0.010}$ & $\bm{0.036}$ \\
     &\begin{footnotesize} Baseline 1\end{footnotesize} & 52.94  & 52.95 & 53.30  & 53.30 & 0.922 & 0.929 & 0.403 & 0.461 \\
     &\begin{footnotesize}Baseline 2\end{footnotesize} &  2.869  & 17.33 & 1.897 & 8.471 & \textbf{--} & \textbf{--} & \textbf{--}& \textbf{--}\\
     &\begin{footnotesize}Baseline 3\end{footnotesize} &  5.966  & 13.43 & 6.377 & 14.11 & 1.145 & 2.136 &1.115 & 2.347 \\
     \hline
        \multicolumn{2}{c|}{\multirow{2}{*}{Settings}} &  \begin{footnotesize}
         $n$=25600
     \end{footnotesize} & \begin{footnotesize}$n$=25600\end{footnotesize}  &\begin{footnotesize} $n$=102400\end{footnotesize} &\begin{footnotesize} $n$=102400 \end{footnotesize} &\begin{footnotesize}$n$=25600 \end{footnotesize} & \begin{footnotesize}$n$=25600 \end{footnotesize} &\begin{footnotesize} $n$=102400\end{footnotesize} & \begin{footnotesize}$n$=102400\end{footnotesize}  \\
     & &\begin{footnotesize} $\sigma = 0.1$\end{footnotesize} & \begin{footnotesize}$\sigma = 0.5$ \end{footnotesize}&\begin{footnotesize} $\sigma = 0.1$\end{footnotesize}& \begin{footnotesize}$\sigma = 0.5$\end{footnotesize}&\begin{footnotesize}$\sigma = 0.1$ \end{footnotesize}&\begin{footnotesize} $\sigma = 0.5$ \end{footnotesize}&\begin{footnotesize} $\sigma = 0.1$\end{footnotesize}&\begin{footnotesize} $\sigma = 0.5$\end{footnotesize} \\
     \hline
     \multirow{4}{*}{\begin{footnotesize}Case 3\end{footnotesize}} & \begin{footnotesize}SemiPDE \end{footnotesize}& $\bm{0.363}$  & $\bm{0.734}$ & $\bm{0.329}$ & $\bm{0.485}$ & $\bm{0.007}$ & $\bm{0.030}$& $\bm{0.004}$ & $\bm{0.015}$ \\
     & \begin{footnotesize}Baseline 1 \end{footnotesize}& 17.07  & 17.08 & 17.90 & 17.57 &7.801  & 7.790 & 7.600 & 7.674 \\
     &\begin{footnotesize}Baseline 2 \end{footnotesize}&   1.876& 7.165 &  1.391& 3.606 & \textbf{--} & \textbf{--} & \textbf{--}& \textbf{--}  \\
     &\begin{footnotesize}Baseline 3 \end{footnotesize}&  14.24  & 16.29 & 14.24 & 16.66 & 2.447 & 1.999 &2.678& 1.703 \\
     \hline
        \multicolumn{2}{c|}{\multirow{2}{*}{Settings}} &  \begin{footnotesize}
         $n$=12800
     \end{footnotesize} & \begin{footnotesize}$n$=12800\end{footnotesize}  &\begin{footnotesize} $n$=102400\end{footnotesize} &\begin{footnotesize} $n$=102400 \end{footnotesize} &\begin{footnotesize}$n$=12800 \end{footnotesize} & \begin{footnotesize}$n$=12800 \end{footnotesize} &\begin{footnotesize} $n$=102400 \end{footnotesize} & \begin{footnotesize}$n$=102400\end{footnotesize}  \\
     & &\begin{footnotesize} $\sigma = 0.1$\end{footnotesize} & \begin{footnotesize}$\sigma = 0.5$ \end{footnotesize}&\begin{footnotesize} $\sigma = 0.1$\end{footnotesize}& \begin{footnotesize}$\sigma = 0.5$\end{footnotesize}&\begin{footnotesize}$\sigma = 0.1$ \end{footnotesize}&\begin{footnotesize} $\sigma = 0.5$ \end{footnotesize}&\begin{footnotesize} $\sigma = 0.1$\end{footnotesize}&\begin{footnotesize} $\sigma = 0.5$\end{footnotesize} \\
     \hline
     \multirow{4}{*}{\begin{footnotesize}Case 4\end{footnotesize}} & \begin{footnotesize}SemiPDE\end{footnotesize} & $\bm{1.572}$  & $\bm{2.843}$ & $\bm{1.472}$ & $\bm{2.009}$ & $\bm{0.723}$ &  $\bm{0.790}$ & $\bm{0.737}$ & $\bm{0.747}$ \\
     & \begin{footnotesize}Baseline 1 \end{footnotesize}& 11.81  & 11.81 & 11.87 &11.87  & 1.813 & 1.829 & 1.905 & 1.871 \\
     &\begin{footnotesize}Baseline 2 \end{footnotesize}&  3.064 &12.96  & 2.561 & 5.164 & \textbf{--} & \textbf{--} & \textbf{--}& \textbf{--} \\
     &\begin{footnotesize}Baseline 3 \end{footnotesize}&  19.45 & 16.70 & 19.36 & 16.70 & 1.507 &  1.653 & 1.500 & 1.666 \\
		 \bottomrule[1pt]
	\end{tabular}
\end{table}

\begin{figure}[!ht]
    \centering
    \subfigure{
    \includegraphics[scale=0.32]{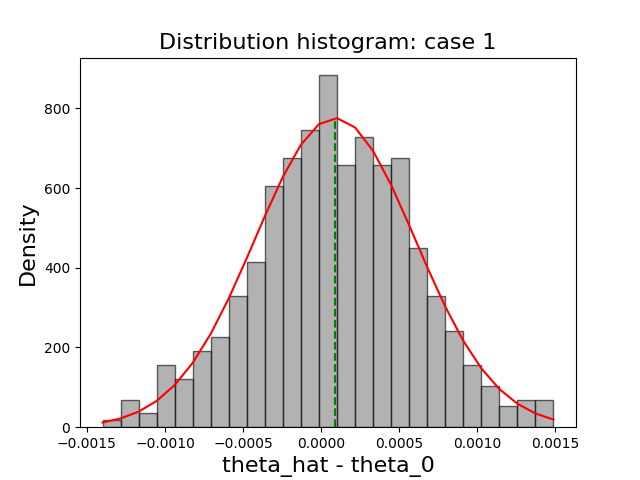}
    \includegraphics[scale=0.32]{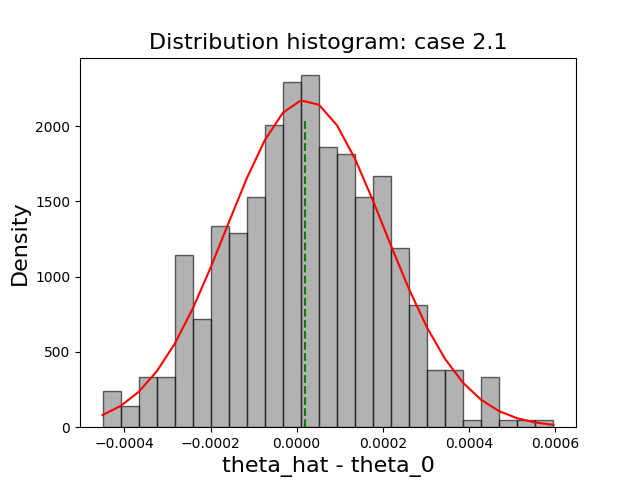}
    \includegraphics[scale=0.32]{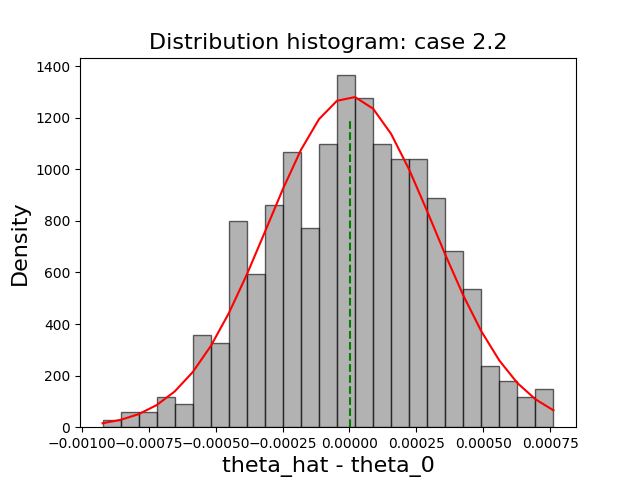}
    }
    \caption{The distribution histogram of $\hat{\theta}_i- \theta_0$. The red line is the density function of normal distribution with mean Bias and standard deviation Std, and the green dash line is Bias.}
    \label{fig:fig_qqplot}
\end{figure}

Second, we investigate the inference performance of  $\hat{\theta}$ for the unknown parameters $\theta$. Due to computational burden, we use Case 1 and Case 2 as representative examples in the following simulations. We set the sample size to $n = 160$ for Case 1 and $n = 640$ for Case 2, with a noise level of $\sigma = 0.1$. 
To estimate the bias and variance of $\hat{\theta}$, we perform $m = 500$ repetitions. The bias and variance are then calculated as
\begin{align*}
    \operatorname{Bias} = \frac{1}{m}\sum_{i=1}^m \hat{\theta}_i - \theta_0,\quad  \operatorname{Std}^2 =\frac{1}{m}\sum_{i=1}^m \left(\hat{\theta}_i - \frac{1}{m}\sum_{i=1}^m \hat{\theta}_i\right)^2, 
\end{align*}
where $\hat{\theta}_i$ is the estimator of $\theta$ in the $i$th experiment. The variance of $\hat{\theta}$ is estimated as $\hat{\sigma}^2 = \gamma^\top\hat{\Sigma}_{\text{eff}}\gamma$ using the estimator $\hat{\Sigma}_{\text{eff}}$ in \eqref{eq:estimation_for_variance}.
Confidence intervals are constructed using Theorem \ref{thm:parametric_inference}. For Case 1, we set $\gamma = 1$, and for Case 2, we use $\gamma = (1,0)^\top,(0,1)^\top$, constructing confidence intervals for $\theta$ in Case 1, and for $\theta_1,\theta_2$ in Case 2. The coverage level of each confidence interval is measured using the coverage rate defined as 
\begin{align*}
    \operatorname{Cover} =\frac{1}{m} \sum_{i=1}^m \mathbf{1}_{\theta_0\in C_{\alpha}(i)},\quad C_{\alpha}(i) = [\hat{\theta}_i - n^{-1/2}z_{1-\alpha/2} \hat{\sigma}_i, \hat{\theta}_i +n^{-1/2}z_{1-\alpha/2} \hat{\sigma}_i],\quad i=1,2,\cdots,m.
\end{align*}
Table \ref{tab:simu:para} presents the bias, variance and coverage rates of the confidence intervals for various significant levels $1-\alpha$. The estimated variance of $\hat{\theta}$ is approximated by $\operatorname{Std}^2$ calculated from repeated experiments. The coverage rate of the constructed confidence interval for $\theta$ meets the significant level $1-\alpha$. 
Figure \ref{fig:fig_qqplot} shows the distribution histogram of $\hat{\theta}_i- \theta_0$, indicating that $\hat{\theta}$ follows an asymptotic normal distribution.


\begin{table}[!ht]
	\centerfloat
     \caption{The estimated bias and standard deviation of the estimator $\hat{\theta}$ and the coverage rates (C.R.) of the constructed confidence intervals at different significant levels, calculated using $500$ repeated random experiments.}
     \label{tab:simu:para}
 \renewcommand\arraystretch{0.8}
 \tabcolsep=0.1cm
 \vspace{0.2cm}
	\begin{tabular}{c|ccccc}
		\toprule[1pt]
     & Bias ($\times 10^{-5}$)& Std ($\times 10^{-4}$)&C.R. of $80\%$ CI & C.R. of $90\%$ CI & C.R. of $95\%$ CI\\
      \midrule[1pt]
      Case 1: $\theta$& $8.857$ & $5.143$&  0.808 & 0.900 & 0.948\\
      Case 2: $\theta_1$ & $2.002$ & $1.834$ &0.794 & 0.906 & 0.952 \\
      Case 2: $\theta_2$ & $0.329$ & $3.113$ & 0.786 & 0.894 & 0.946 \\
		 \bottomrule[1pt]
	\end{tabular}
\end{table}
    


Finally, we examine the identifiability and robustness of the proposed SemiPDE framework. In Cases 1 and 2, we assume known  mechanism, i.e., $\mathcal{F}(u) = 0$, meaning the true model is located within the parametric model family $\mathcal{L}(u,\theta) = 0$. We compare the SemiPDE method with parametric baseline 1 and PINNs baseline 3. All other settings are consistent with those in the first simulation. Table \ref{tab:compared_baseline_paramodel} presents the parameter estimation errors for $\theta$. The results indicate that the SemiPDE method demonstrates minimal efficiency loss compared to the parametric model, even when the true model does not contain an unknown mechanism.
\begin{table}[!ht]
	\centerfloat
     \caption{Parametric estimation errors ($\times 10^{-3}$) for SemiPDE and Baselines with no unknown mechanism.}
	\label{tab:compared_baseline_paramodel}
	\begin{tabular}{c|cccc|cccc}
		\toprule[1pt]
		   &\multicolumn{4}{c|}{ Case 1}&  \multicolumn{4}{c}{Case 2} \\
     \hline 
       &  \begin{footnotesize}
         $n=800$
     \end{footnotesize} & \begin{footnotesize}$n=800$\end{footnotesize}  &\begin{footnotesize} $n=1600$ \end{footnotesize} &\begin{footnotesize} $n=1600$ \end{footnotesize} &\begin{footnotesize}$n=800$ \end{footnotesize} & \begin{footnotesize}$n=800$ \end{footnotesize} &\begin{footnotesize} $n=1600$ \end{footnotesize} & \begin{footnotesize}$n=1600$\end{footnotesize}  \\
     &\begin{footnotesize} $\sigma = 0.1$\end{footnotesize} & \begin{footnotesize}$\sigma = 0.5$ \end{footnotesize}&\begin{footnotesize} $\sigma = 0.1$\end{footnotesize}& \begin{footnotesize}$\sigma = 0.5$\end{footnotesize}&\begin{footnotesize}$\sigma = 0.1$ \end{footnotesize}&\begin{footnotesize} $\sigma = 0.5$ \end{footnotesize}&\begin{footnotesize} $\sigma = 0.1$\end{footnotesize}&\begin{footnotesize} $\sigma = 0.5$\end{footnotesize} \\
     \midrule[1pt]
     SemiPDE & 0.314  & 1.229 &0.450  & 0.957 & 0.264 &  1.023& 0.101 & 0.434 \\
     Baseline 1 & 0.307 & 1.401 & 0.198 &  0.995& 0.202 & 0.957 & 0.080 &  0.401\\
     Baseline 3 & 2.824  & 5.289 & 2.331 & 5.608 & 5.446& 5.645 & 5.184 &5.885  \\
		 \bottomrule[1pt]
	\end{tabular}
    \vspace{-1cm}
\end{table}

\section{Real Data Applications}
\label{sec:realdata}
In this section, we apply the SemiPDE framework to two real-world examples: the reaction-diffusion model in mathematical biology and wave propagation dynamics through vegetation described by the Navier-Stokes equations.

\subsection{The Behaviour of In Vitro Cell Culture Assays}

In cell biology, in vitro cell culture assays are commonly used to study the behavior of cell populations in various environments. A widely observed phenomenon in these assays is classical reaction-diffusion dynamics, exemplified by scratch assays. We utilized experimental data from \cite{jin2016reproducibility}, which included a comprehensive dataset obtained through scratch assays using the PC-3 prostate cancer cell line and can be downloaded at \href{http://dx.doi.org/10.1016/j.jtbi.2015.10.040}{http://dx.doi.org/10.1016/j.jtbi.2015.10.040}. In the dataset, cell densities were collected under six different environmental settings across three repeated experiments. Observations $Y(t,x)$ were recorded five times $t_i$ over a 48-hour period, with data from 38 consistent monolayers $x_i$ collected at each time. 


Previous studies \citep{liang2007vitro, jin2016reproducibility} have employed two types of partial differential equations (PDEs) to model the dynamics of cell densities $u(t,x)$ in scratch assays, referred to here as Benchmark 1 and Benchmark 2. The first model is the Fisher-Kolmogorov equation and the second is the Porous-Fisher equation:
\begin{align*}
    \text{Benchmark 1:}\quad &\partial_t u(t,x) - \theta\Delta u(t,x) - \lambda u(t,x)\left(1-\frac{u(t,x)}{K}\right) = 0\\
     \text{Benchmark 2:}\quad &\partial_t u(t,x) - \theta\partial_x\left(\frac{u(t,x)}{K}\partial_x u(t,x)\right) -  \lambda u(t,x)\left(1-\frac{u(t,x)}{K}\right) = 0,
\end{align*}
where $\theta$ is the unknown cell diffusivity, $\lambda$ is the
cell proliferation rate and $K$ is the maximum cell density
for a monolayer. Neumann boundary condition $\partial_x u = 0$ and a known initial condition $u(0,x)$ are applied. 

\begin{figure}[!ht]
\makeatletter\def\@captype{table}\makeatother
\begin{minipage}{0.54\linewidth}
\centerfloat
	\begin{tabular}{c|cc}
		\toprule[1pt]
		Model &$\theta$ &MSE   \\
  \midrule[1pt]
  SemiPDE 1 & 1232 & $1.638\times 10^{-8}$ \\
  Benchmark 1 & 853 & $1.874\times 10^{-8}$ \\
  SemiPDE 2 & 897 & $1.769\times 10^{-8}$ \\
  Benchmark 2 & 833 & $1.840\times 10^{-8}$ \\
  \bottomrule[1pt]
	\end{tabular}
      \caption{Estimated $\theta$ and MSE for all SemiPDE models and benchmarks.}
    \label{tab:table_realdata1}
\end{minipage}  
\makeatletter\def\@captype{figure}\makeatother
\begin{minipage}{0.45\linewidth}
        \centering
		\includegraphics[width=0.99\columnwidth]{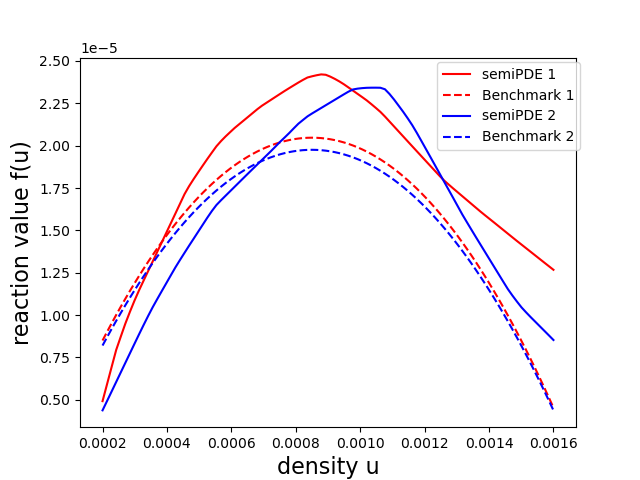}
        \vspace{-1cm}
		\caption{The estimated reaction mechanism $f(u)$ for each model.}
		\label{fig:fig_realdata1}
\end{minipage}
\end{figure}

As discussed in Example \ref{emp:r-d}, two critical aspects of the reaction-diffusion dynamics are: 1) the unknown cell diffusivity $\theta$, and 2) the local reaction mechanism, represented as a function of the current cell density $u(t,x)$, i.e., $f(u)$. We construct SemiPDE models that integrate the known evolution and diffusion mechanisms, or their variations, with the unknown reaction mechanism. These models are identified as SemiPDE 1 and SemiPDE 2 respectively:
\vspace{-0.25cm}
\begin{align*}
     \text{SemiPDE 1:} \quad & \partial_t u(t,x) - \theta\Delta u(t,x) -f(u(t,x)) = 0\\
     \text{SemiPDE 2:} \quad & \partial_t u(t,x) - \theta\partial_x\left(\frac{u(t,x)}{K}\partial_x u(t,x)\right) -f(u(t,x))= 0.
\end{align*}
Data from the three repeated experiments across six different settings are used. Of these, 15 experiments are randomly selected as training data while the remaining three are reserved for testing. The mean square error (MSE) is calculated on the testing data as $\sum_{i=1}^n (Y_i-\tilde{u}(t_i,x_i))^2/n$ where $\tilde{u}(\cdot)$ represents the numerical solution for each model. The results for all SemiPDE models and benchmarks are presented in Table \ref{tab:table_realdata1} and profiles of $f(u)$ are compared in Figure \ref{fig:fig_realdata1}. The MSE on the testing data is lower for the SemiPDE models than for the benchmark models, indicating that the SemiPDE models capture the underlying data mechanism more effectively.

The estimates of $\theta$ in our two SemiPDE models are significantly larger than those from the benchmark models, yet they still fall within the range reported in \cite{liang2007vitro}. Regarding the unknown mechanism, the reaction term $f(u)$ estimated in our SemiPDE models initially increases as the density $u$ grows from zero, but then decreases as $u$ approaches to the maximum density $K$. 
However, our estimated reaction mechanism function exhibits lower values as $u$ approaches zero and higher values as $u$ nears maximum density, compared to benchmark reaction terms. 
This suggests that cell proliferation rates are more influenced by low cell numbers than by proximity to the upper limit of cell density.
This phenomenon may result from the large cell diffusivity $\theta$ in the diffusion term of our SemiPDE models, which enhances cell transport to adjacent monolayers, thereby mitigating the inhibitory effect of high density. According to \cite{shi2006persistence,wang2019persistence}, the asymmetry of the reaction term $f(u)$ is also attributed to the Allee effect in population biology.

\subsection{Wave Propagation Through Vegetation}

Wave transformation and dissipation in coastal areas are typical scenarios where Navier-Stokes equations, as described in Example \ref{emp: n-s}, are applied. These studies often focus on how varying coastal environments influence water waves. We use the laboratory data from \cite{hu2021laboratory}, which included experiments on wave dissipation process in vegetation under both emergent and submerged conditions with underlying currents. The dataset can be downloaded at \href{https://doi.org/10.6084/m9.figshare.13026530.v2}{https://doi.org/10.6084/m9.figshare.13026530.v2}. 
The dataset comprised horizon velocity $u(t,x)$ from 668 tests under different environment settings. We specifically analyze the 2019 experiments in this dataset, with wave height $0.03m$, wave period $0.8 s$ and imposed currents $+0.03 m/s$. Four cases were considered, involving two water depths $h=0.2m, 0.33 m$ and two mimic stem densities $N_v=139 \operatorname{stem}/m^2, 556 \operatorname{stem}/m^2$, with each case repeated three times.

To construct the SemiPDE model, we employ the one-dimensional Navier-Stokes equation proposed in \cite{marmanis2006one} and consider the force term expressed as the sum of drag force and inertia force as the known mechanism. We then apply a data-driven unknown mechanism $f(t,x)$ involving the velocity $v,w$ and pressure $p$. Our SemiPDE model is then 
\begin{align*}
    \left(1+\frac{\pi}{4}\rho C_M d^2_v N_v\right)\partial_t u + u \nabla u - \nu \Delta u + \frac{1}{2}\rho C_D d_v N_v u|u| - f(t,x) = 0.
\end{align*}
The parameters in this PDE are $\theta = (C_D,\nu)$, where $\nu$ is a parameter similar to viscosity and $C_D$ is the vegetation drag coefficient we are interested in. 

Three benchmarks for estimating $C_D$ are included: Benchmark 1 uses the least square method in the force term with extra force data; Benchmark 2 employs a direct measurement method with extra velocity and force data; and Benchmark 3 utilizes the parametric PDE model by using the known mechanism mentioned before.

\begin{figure}[!ht]
\vspace{0.6cm}
\makeatletter\def\@captype{table}\makeatother
\begin{minipage}{0.54\linewidth}
\centerfloat
	\begin{tabular}{c|cccc}
		\toprule[1pt]
		Method & Case 1 & Case 2& Case 3 & Case 4\\
  \midrule[1pt]
  SemiPDE &2.0882&1.6243&2.2224&2.6844 \\
  \hline
  Benchmark 1&2.1756&1.8689&2.5167&2.7778\\
  \hline
  Benchmark 2&2.4906&2.0845&1.8613&3.1641\\
  \hline
  Benchmark 3& 5.5839&0.7794&1.4185&0.4297\\
  \bottomrule[1pt]
	\end{tabular}
      \caption{Estimated value of $C_D$ for our SemiPDE model and benchmarks.}
    \label{tab:table_realdata2}
\end{minipage}  
\makeatletter\def\@captype{figure}\makeatother
\begin{minipage}{0.45\linewidth}
        \centering
		\includegraphics[width=0.99\columnwidth]{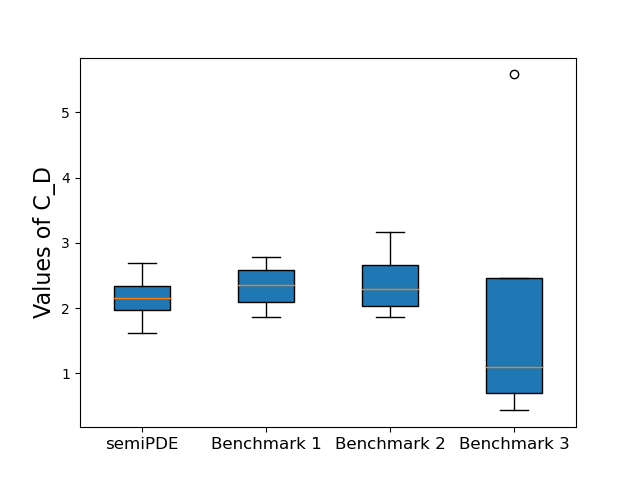}
        \vspace{-1cm}
		\caption{Boxplot for estimated $C_D$ in different environments.}
		\label{fig: boxplot_Cd}
\end{minipage}
\end{figure}

By applying our SemiPDE framework, the estimated values of $C_D$ are presented in Table \ref{tab:table_realdata2} and Figure \ref{fig: boxplot_Cd}, alongside comparisons with those three benchmarks. 
Benchmark 3, which employs a parametric PDE model, suffers significantly from model misspecification due to the complexity of real-world environments, leading to poor estimates of $C_D$. In contrast, our model achieves higher accuracy, aided by mechanisms that remain unknown. Compared to Benchmarks 1 and 2, our SemiPDE model yields more stable estimation of  $C_D$ and shows minimal variation across different experimental conditions. This empirical evidence demonstrates the advantage of the proposed SemiPDE approach which 
leverages known mechanisms so that no additional data such as  force sensor readings or height measurements are required.

\section{Conclusion and Discussion}
\label{sec:discussion}
In this paper, we introduce the SemiPDE framework to model, estimate and infer the generation and evolution processes of data commonly observed in scientific research.
The SemiPDE framework integrates semiparametric statistical methods with modern deep learning techniques in PDE models. This combination provides new statistical estimation and inference tools for a wide range of practical applications. Future research may involve applying specific SemiPDE models to real-world problems and verifying the critical though challenging assumptions, such as Lipschitz and stability conditions. The framework can also be extended to handle unknown or complex initial and boundary conditions, allowing for more realistic modeling scenarios. Furthermore, incorporating repeated measurements, sparse observations and online duration data can further improve the SemiPDE modeling of realistic data generation and evolution. Overall, the proposed SemiPDE framework offers an interpretable and flexible approach to analyzing data with underlying mechanisms and shows strong potential for broader applications.



\section{Acknowledgments}
Ziyuan Chen is the first author. Fang Yao is the corresponding author. This research is partially supported by the National Key Research and Development Program of China (No.
2022YFA1003801), the National Natural Science Foundation of China (No. 12292981, 11931001, 123B2009), the New Cornerstone Science Foundation through the Xplorer Prize, the LMAM, the Fundamental Research Funds for the Central Universities, Peking University, and LMEQF.

\bibliographystyle{abbrvnat}
\bibliography{reference}

\begin{thebibliography}{47}
\providecommand{\natexlab}[1]{#1}
\providecommand{\url}[1]{\texttt{#1}}
\expandafter\ifx\csname urlstyle\endcsname\relax
  \providecommand{\doi}[1]{doi: #1}\else
  \providecommand{\doi}{doi: \begingroup \urlstyle{rm}\Url}\fi

\bibitem[Berg and Nystr{\"o}m(2019)]{berg2019data}
J.~Berg and K.~Nystr{\"o}m.
\newblock Data-driven discovery of pdes in complex datasets.
\newblock \emph{Journal of Computational Physics}, 384:\penalty0 239--252, 2019.

\bibitem[Blazek(2015)]{blazek2015computational}
J.~Blazek.
\newblock \emph{Computational fluid dynamics: principles and applications}.
\newblock Butterworth-Heinemann, 2015.

\bibitem[Cao et~al.(2012)Cao, Huang, and Wu]{cao2012penalized}
J.~Cao, J.~Z. Huang, and H.~Wu.
\newblock Penalized nonlinear least squares estimation of time-varying parameters in ordinary differential equations.
\newblock \emph{Journal of computational and graphical statistics}, 21\penalty0 (1):\penalty0 42--56, 2012.

\bibitem[Chernozhukov et~al.(2018)Chernozhukov, Chetverikov, Demirer, Duflo, Hansen, Newey, and Robins]{chernozhukov2018double}
V.~Chernozhukov, D.~Chetverikov, M.~Demirer, E.~Duflo, C.~Hansen, W.~Newey, and J.~Robins.
\newblock Double/debiased machine learning for treatment and structural parameters, 2018.

\bibitem[Dai et~al.(2016)Dai, Tong, and Genton]{dai2016optimal}
W.~Dai, T.~Tong, and M.~G. Genton.
\newblock Optimal estimation of derivatives in nonparametric regression.
\newblock \emph{Journal of Machine Learning Research}, 17\penalty0 (164):\penalty0 1--25, 2016.

\bibitem[Dai and Li(2022)]{dai2022kernel}
X.~Dai and L.~Li.
\newblock Kernel ordinary differential equations.
\newblock \emph{Journal of the American Statistical Association}, 117\penalty0 (540):\penalty0 1711--1725, 2022.

\bibitem[De~Brabanter et~al.(2013)De~Brabanter, De~Brabanter, De~Moor, and Gijbels]{de2013derivative}
K.~De~Brabanter, J.~De~Brabanter, B.~De~Moor, and I.~Gijbels.
\newblock Derivative estimation with local polynomial fitting.
\newblock \emph{Journal of Machine Learning Research}, 14\penalty0 (1), 2013.

\bibitem[De~Ryck and Mishra(2024)]{de2024numerical}
T.~De~Ryck and S.~Mishra.
\newblock Numerical analysis of physics-informed neural networks and related models in physics-informed machine learning.
\newblock \emph{Acta Numerica}, 33:\penalty0 633--713, 2024.

\bibitem[Du et~al.(2024)Du, Chen, and Zhang]{du2024discover}
M.~Du, Y.~Chen, and D.~Zhang.
\newblock Discover: Deep identification of symbolically concise open-form partial differential equations via enhanced reinforcement learning.
\newblock \emph{Physical Review Research}, 6\penalty0 (1):\penalty0 013182, 2024.

\bibitem[Giles and Pierce(2000)]{giles2000introduction}
M.~B. Giles and N.~A. Pierce.
\newblock An introduction to the adjoint approach to design.
\newblock \emph{Flow, turbulence and combustion}, 65:\penalty0 393--415, 2000.

\bibitem[Ho et~al.(1995)Ho, Neumann, Perelson, Chen, Leonard, and Markowitz]{ho1995rapid}
D.~D. Ho, A.~U. Neumann, A.~S. Perelson, W.~Chen, J.~M. Leonard, and M.~Markowitz.
\newblock Rapid turnover of plasma virions and cd4 lymphocytes in hiv-1 infection.
\newblock \emph{Nature}, 373\penalty0 (6510):\penalty0 123--126, 1995.

\bibitem[Hu et~al.(2021)Hu, Lian, Wei, Li, Stive, and Suzuki]{hu2021laboratory}
Z.~Hu, S.~Lian, H.~Wei, Y.~Li, M.~Stive, and T.~Suzuki.
\newblock Laboratory data on wave propagation through vegetation with following and opposing currents.
\newblock \emph{Earth System Science Data}, 13\penalty0 (10):\penalty0 4987--4999, 2021.

\bibitem[Isakov(2006)]{isakov2006inverse}
V.~Isakov.
\newblock \emph{Inverse problems for partial differential equations}, volume 127.
\newblock Springer, 2006.

\bibitem[Jeng et~al.(2013)Jeng, Ye, Zhang, and Liu]{jeng2013integrated}
D.-S. Jeng, J.-H. Ye, J.-S. Zhang, and P.-F. Liu.
\newblock An integrated model for the wave-induced seabed response around marine structures: Model verifications and applications.
\newblock \emph{Coastal Engineering}, 72:\penalty0 1--19, 2013.

\bibitem[Jia et~al.(2023)Jia, Wu, Li, and Meng]{jia2023variational}
J.~Jia, Y.~Wu, P.~Li, and D.~Meng.
\newblock Variational inverting network for statistical inverse problems of partial differential equations.
\newblock \emph{Journal of Machine Learning Research}, 24\penalty0 (201):\penalty0 1--60, 2023.

\bibitem[Jin et~al.(2016)Jin, Shah, Penington, McCue, Chopin, and Simpson]{jin2016reproducibility}
W.~Jin, E.~T. Shah, C.~J. Penington, S.~W. McCue, L.~K. Chopin, and M.~J. Simpson.
\newblock Reproducibility of scratch assays is affected by the initial degree of confluence: experiments, modelling and model selection.
\newblock \emph{Journal of theoretical biology}, 390:\penalty0 136--145, 2016.

\bibitem[Johnson(2009)]{johnson2009numerical}
C.~Johnson.
\newblock \emph{Numerical solution of partial differential equations by the finite element method}.
\newblock Courier Corporation, 2009.

\bibitem[Kaltenbacher and Rundell(2020)]{kaltenbacher2020inverse}
B.~Kaltenbacher and W.~Rundell.
\newblock The inverse problem of reconstructing reaction--diffusion systems.
\newblock \emph{Inverse Problems}, 36\penalty0 (6):\penalty0 065011, 2020.

\bibitem[Liang et~al.(2007)Liang, Park, and Guan]{liang2007vitro}
C.-C. Liang, A.~Y. Park, and J.-L. Guan.
\newblock In vitro scratch assay: a convenient and inexpensive method for analysis of cell migration in vitro.
\newblock \emph{Nature protocols}, 2\penalty0 (2):\penalty0 329--333, 2007.

\bibitem[Liang and Wu(2008)]{liang2008parameter}
H.~Liang and H.~Wu.
\newblock Parameter estimation for differential equation models using a framework of measurement error in regression models.
\newblock \emph{Journal of the American Statistical Association}, 103\penalty0 (484):\penalty0 1570--1583, 2008.

\bibitem[Lu et~al.(2021)Lu, Meng, Mao, and Karniadakis]{lu2021deepxde}
L.~Lu, X.~Meng, Z.~Mao, and G.~E. Karniadakis.
\newblock Deepxde: A deep learning library for solving differential equations.
\newblock \emph{SIAM review}, 63\penalty0 (1):\penalty0 208--228, 2021.

\bibitem[Marmanis et~al.(2006)Marmanis, Hamman, and Kirby]{marmanis2006one}
H.~Marmanis, C.~Hamman, and R.~Kirby.
\newblock A one-dimensional model of the navier-stokes.
\newblock Technical report, Citeseer, 2006.

\bibitem[Nickl(2017)]{nickl2017bayesian}
R.~Nickl.
\newblock On bayesian inference for some statistical inverse problems with partial differential equations.
\newblock \emph{Bernoulli News}, 24\penalty0 (2):\penalty0 5--9, 2017.

\bibitem[Nickl(2023)]{nickl2023bayesian}
R.~Nickl.
\newblock Bayesian non-linear statistical inverse problems.
\newblock 2023.

\bibitem[Pantano et~al.(2003)Pantano, Boyce, and Parks]{pantano2003nonlinear}
A.~Pantano, M.~C. Boyce, and D.~M. Parks.
\newblock Nonlinear structural mechanics based modeling of carbon nanotube deformation.
\newblock \emph{Physical review letters}, 91\penalty0 (14):\penalty0 145504, 2003.

\bibitem[Podina et~al.(2023)Podina, Eastman, and Kohandel]{podina2023universal}
L.~Podina, B.~Eastman, and M.~Kohandel.
\newblock Universal physics-informed neural networks: symbolic differential operator discovery with sparse data.
\newblock In \emph{International Conference on Machine Learning}, pages 27948--27956. PMLR, 2023.

\bibitem[Rackauckas et~al.(2020)Rackauckas, Ma, Martensen, Warner, Zubov, Supekar, Skinner, Ramadhan, and Edelman]{rackauckas2020universal}
C.~Rackauckas, Y.~Ma, J.~Martensen, C.~Warner, K.~Zubov, R.~Supekar, D.~Skinner, A.~Ramadhan, and A.~Edelman.
\newblock Universal differential equations for scientific machine learning.
\newblock \emph{arXiv preprint arXiv:2001.04385}, 2020.

\bibitem[Raissi et~al.(2019)Raissi, Perdikaris, and Karniadakis]{raissi2019physics}
M.~Raissi, P.~Perdikaris, and G.~E. Karniadakis.
\newblock Physics-informed neural networks: A deep learning framework for solving forward and inverse problems involving nonlinear partial differential equations.
\newblock \emph{Journal of Computational physics}, 378:\penalty0 686--707, 2019.

\bibitem[Ramsay et~al.(2007)Ramsay, Hooker, Campbell, and Cao]{ramsay2007parameter}
J.~O. Ramsay, G.~Hooker, D.~Campbell, and J.~Cao.
\newblock Parameter estimation for differential equations: a generalized smoothing approach.
\newblock \emph{Journal of the Royal Statistical Society Series B: Statistical Methodology}, 69\penalty0 (5):\penalty0 741--796, 2007.

\bibitem[Rasmussen(2012)]{rasmussen2012dynamic}
B.~P. Rasmussen.
\newblock Dynamic modeling for vapor compression systems—part i: Literature review.
\newblock \emph{HVAC\&R Research}, 18\penalty0 (5):\penalty0 934--955, 2012.

\bibitem[Sakthivel et~al.(2011)Sakthivel, Gnanavel, Balan, and Balachandran]{sakthivel2011inverse}
K.~Sakthivel, S.~Gnanavel, N.~B. Balan, and K.~Balachandran.
\newblock Inverse problem for the reaction diffusion system by optimization method.
\newblock \emph{Applied mathematical modelling}, 35\penalty0 (1):\penalty0 571--579, 2011.

\bibitem[Schmidt-Hieber(2020)]{schmidt2020nonparametric}
J.~Schmidt-Hieber.
\newblock Nonparametric regression using deep neural networks with relu activation function.
\newblock \emph{The Annals of Statistics}, 48\penalty0 (4):\penalty0 1875, 2020.

\bibitem[Sgura et~al.(2019)Sgura, Lawless, and Bozzini]{sgura2019parameter}
I.~Sgura, A.~S. Lawless, and B.~Bozzini.
\newblock Parameter estimation for a morphochemical reaction-diffusion model of electrochemical pattern formation.
\newblock \emph{Inverse Problems in Science and Engineering}, 27\penalty0 (5):\penalty0 618--647, 2019.

\bibitem[Shi and Shivaji(2006)]{shi2006persistence}
J.~Shi and R.~Shivaji.
\newblock Persistence in reaction diffusion models with weak allee effect.
\newblock \emph{Journal of Mathematical Biology}, 52\penalty0 (6):\penalty0 807--829, 2006.

\bibitem[Singh and Saha~Ray(2022)]{singh2022analysis}
S.~Singh and S.~Saha~Ray.
\newblock Analysis of stochastic fitzhugh--nagumo equation for wave propagation in a neuron arising in certain neurobiology models.
\newblock \emph{International Journal of Biomathematics}, 15\penalty0 (05):\penalty0 2250027, 2022.

\bibitem[Stockie(2011)]{stockie2011mathematics}
J.~M. Stockie.
\newblock The mathematics of atmospheric dispersion modeling.
\newblock \emph{Siam Review}, 53\penalty0 (2):\penalty0 349--372, 2011.

\bibitem[Stone(1980)]{stone1980optimal}
C.~J. Stone.
\newblock Optimal rates of convergence for nonparametric estimators.
\newblock \emph{The annals of Statistics}, pages 1348--1360, 1980.

\bibitem[Tan et~al.(2024)Tan, Zhang, Wang, Huang, and Yao]{tan2024green}
J.~Tan, G.~Zhang, X.~Wang, H.~Huang, and F.~Yao.
\newblock Green’s matching: an efficient approach to parameter estimation in complex dynamic systems.
\newblock \emph{Journal of the Royal Statistical Society Series B: Statistical Methodology}, page qkae031, 2024.

\bibitem[Tenachi et~al.(2023)Tenachi, Ibata, and Diakogiannis]{tenachi2023deep}
W.~Tenachi, R.~Ibata, and F.~I. Diakogiannis.
\newblock Deep symbolic regression for physics guided by units constraints: toward the automated discovery of physical laws.
\newblock \emph{The Astrophysical Journal}, 959\penalty0 (2):\penalty0 99, 2023.

\bibitem[Vadeboncoeur et~al.(2023)Vadeboncoeur, Akyildiz, Kazlauskaite, Girolami, and Cirak]{vadeboncoeur2023fully}
A.~Vadeboncoeur, {\"O}.~D. Akyildiz, I.~Kazlauskaite, M.~Girolami, and F.~Cirak.
\newblock Fully probabilistic deep models for forward and inverse problems in parametric pdes.
\newblock \emph{Journal of Computational Physics}, 491:\penalty0 112369, 2023.

\bibitem[Wang et~al.(2014)Wang, Cao, Ramsay, Burger, Laporte, and Rockstroh]{wang2014estimating}
L.~Wang, J.~Cao, J.~O. Ramsay, D.~Burger, C.~Laporte, and J.~K. Rockstroh.
\newblock Estimating mixed-effects differential equation models.
\newblock \emph{Statistics and Computing}, 24:\penalty0 111--121, 2014.

\bibitem[Wang et~al.(2019)Wang, Shi, and Wang]{wang2019persistence}
Y.~Wang, J.~Shi, and J.~Wang.
\newblock Persistence and extinction of population in reaction--diffusion--advection model with strong allee effect growth.
\newblock \emph{Journal of mathematical biology}, 78:\penalty0 2093--2140, 2019.

\bibitem[Wu and Ding(1999)]{wu1999population}
H.~Wu and A.~A. Ding.
\newblock Population hiv-1 dynamics in vivo: applicable models and inferential tools for virological data from aids clinical trials.
\newblock \emph{Biometrics}, 55\penalty0 (2):\penalty0 410--418, 1999.

\bibitem[Xun et~al.(2013)Xun, Cao, Mallick, Maity, and Carroll]{xun2013parameter}
X.~Xun, J.~Cao, B.~Mallick, A.~Maity, and R.~J. Carroll.
\newblock Parameter estimation of partial differential equation models.
\newblock \emph{Journal of the American Statistical Association}, 108\penalty0 (503):\penalty0 1009--1020, 2013.

\bibitem[Yan et~al.(2025)Yan, Chen, and Yao]{yan2025semiparametric}
S.~Yan, Z.~Chen, and F.~Yao.
\newblock Semiparametric m-estimation with overparameterized neural networks.
\newblock \emph{arXiv preprint arXiv:2504.19089}, 2025.

\bibitem[Yang et~al.(2021)Yang, Wong, and Kou]{yang2021inference}
S.~Yang, S.~W. Wong, and S.~Kou.
\newblock Inference of dynamic systems from noisy and sparse data via manifold-constrained gaussian processes.
\newblock \emph{Proceedings of the National Academy of Sciences}, 118\penalty0 (15):\penalty0 e2020397118, 2021.

\bibitem[Yin et~al.(2021)Yin, Le~Guen, Dona, Ayed, de~B{\'e}zenac, Thome, and Gallinari]{yin2021augmenting}
Y.~Yin, V.~Le~Guen, J.~Dona, I.~Ayed, E.~de~B{\'e}zenac, N.~Thome, and P.~Gallinari.
\newblock Augmenting physical models with deep networks for complex dynamics forecasting.
\newblock In \emph{Ninth International Conference on Learning Representations ICLR 2021}, 2021.

\end{thebibliography}

\end{document}